\pdfoutput=1
\documentclass{acm_proc_article-sp}
\usepackage{graphicx}
\usepackage{amsmath}
\usepackage{multirow}
\usepackage{booktabs}
\usepackage{subfigure}
\usepackage{graphics}
\usepackage[ruled]{algorithm2e}
\renewcommand{\algorithmcfname}{ALGORITHM}

\begin{document}

\title{Modeling Movements in Oil, Gold, Forex and Market Indices using Search Volume Index and Twitter Sentiments}

\numberofauthors{2}
\author{
\alignauthor
Tushar Rao\\
       \affaddr{Netaji Subhas Institute of Technology}\\
       \affaddr{Delhi, India}\\
       \email{rao.tushar@nsitonline.in}
\alignauthor
Saket Srivastava\\
       \affaddr{IIITD}\\
       \affaddr{Delhi, India}\\
       \email{saket@iiitd.ac.in}
}
\maketitle
\vspace{-0.3cm}
\begin{abstract}
 Study of the forecasting models using large scale microblog discussions and the search behavior data can provide a good insight for better understanding the market movements. In this work we collected a dataset of 2 million tweets and search volume index (SVI from Google) for a period of June 2010 to September 2011. We perform a study over a set of comprehensive causative relationships and developed a unified approach to a model for various market securities like equity (Dow Jones Industrial Average-DJIA and NASDAQ-100), commodity markets (oil and gold) and Euro Forex rates. We also investigate the lagged and statistically causative relations of Twitter sentiments developed during active trading days and market inactive days in combination with the search behavior of public before any change in the prices/ indices.
Our results show extent of lagged significance with high correlation value upto 0.82 between search volumes and gold price in USD. We find weekly accuracy in direction (up and down prediction) uptil 94.3\% for DJIA and 90\% for NASDAQ-$100$ with significant reduction in mean average percentage error for all the forecasting models.
\end{abstract}
\vspace{-0.2cm}
%

\vspace{-0.15cm}

\section{INTRODUCTION}
\label{intro}
Most of the earlier works in computational finance comprise of \textit{efficient market hypothesis} (EMH) that asserts market movements at the present level are a function of already existing news, whispers and the future valuation of dividends of a stock/ company \cite{Guresen201110389, Hong}. However research by Qian et al. shows markets are not fully efficient \cite{Qian}. Behavioral finance is attracted high interest by financial \cite{kahneman} community. It challenges the very existence of efficient markets by placing the role of human sentiment and the social mood as vital part of investment decisions \cite{Nofsinger}. It challenges the EMH by adding the notion of human emotion and the macro-level mood play into investment decisions. For example at micro-level, consistently rising stocks is an indication of \emph{selling to hold} the profits and perform subsequent portfolio adjustments. However surprising index trends are observed at macro-economic level. Another example of positive stock sentiment resulting in negative price movement is observed when if lots of people exude high confidence while making buy decision for a commodity or stock, causing the price to rise so steeply that instead of stabilizing it falls. To further elaborate this point we will discuss the results about how bullish twitter sentiments can yield negative correlations with DJIA index (in other word bearish).
\vspace{-0.15cm}

This era of web technology is marked with high entropy of information spread as well as retrieval~\cite{danah_article}. Earlier in late 90s before the spread of social web, information regarding commodities/currency rates and buy/sell sentiments took a long time (maybe even full day) to disseminate fully in the investor community. Also, the companies and markets took a long time (weeks or months) to calm market rumors, news or false information. This provides an opportunity to researchers to develop web mining platforms targeted towards mining relevant financial insights from the social media and web.

\vspace{-0.2cm}
In social web mining context, distinctively there are two different approaches that researchers have taken for market prediction. Firstly social media feeds (such as Twitter API) can provide an important resource to measure investor mood at comprehensive scale \cite{Ant_frank,Gilbert_Karahalios_2010,Sprenger,Bollen_Mao_Zeng_2010, Bollen_second_paper, facebook_stock}. Secondly search volumes (eg. using Google search) related to financial market instruments (stocks, bonds, indices, commodities etc.) have been shown to give out predictive and causative relationships with the market returns \cite{tetlock}.

\begin {figure*}[htpll]
\label{fig:flow_chart}
\vspace{-0.2in}
\includegraphics [scale=0.375] {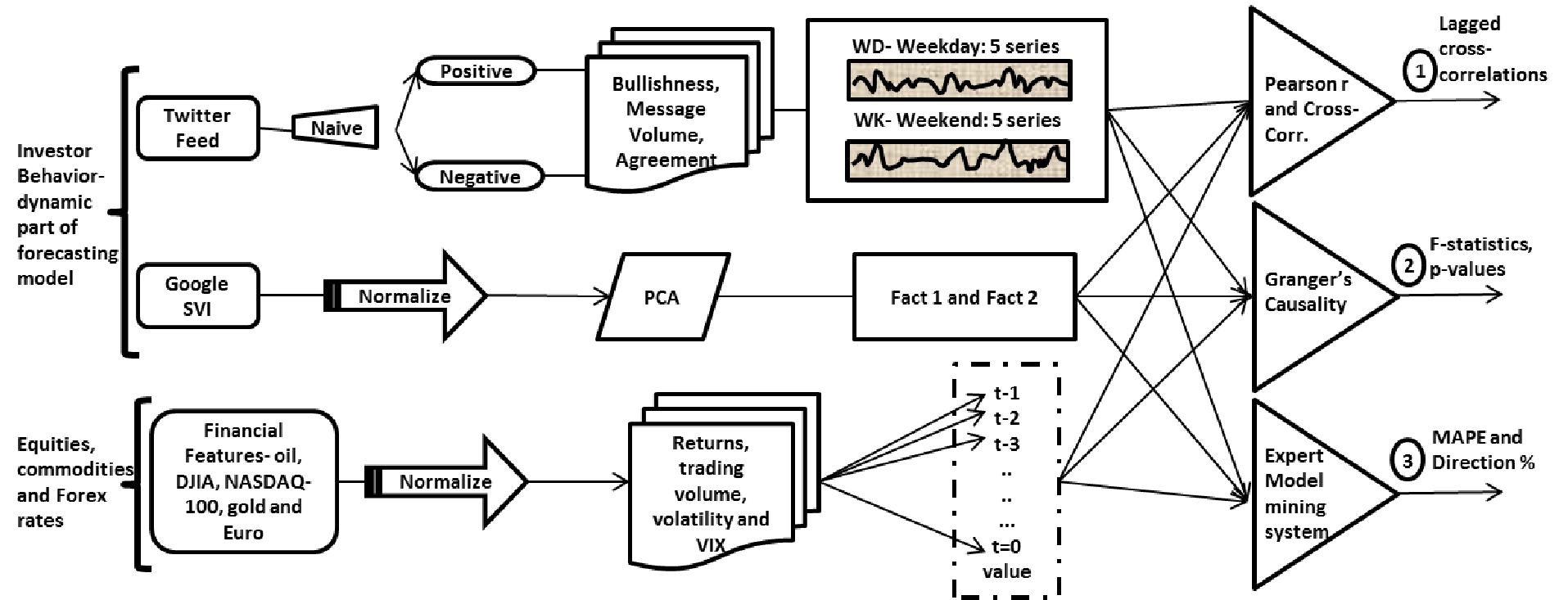}
\vspace{-0.2cm}
\caption {Flowchart of the proposed methodology showing the various phases
of sentimental analysis beginning with SVI/ Tweet collection to stock future prediction.
In the final phase three set of results have been presented:(1) Correlation
results for twitter sentiments and stock prices for different companies
(2) Granger's casuality analysis to prove that the stock prices are affected in
 the short term by Twitter sentiments (3) Using EMMS for quantitative
 comparison in stock market prediction using tweet features}
\end{figure*}

\vspace{-0.2cm}
Bollen et al. has used dimensions of Google- Profile of Mood
States to reflect changes in closing price of DJIA \cite{Bollen_Mao_Zeng_2010}. Another work by Mao et al. covers effect of search volumes data in description with the preliminary sentiment indices of entire twitter feed on stock market movements of DJIA and volatility index of commodities like gold \cite{Bollen_2nd_paper}.
Zhang et al. also made have made use of dimensions in
human behavior- fear and hope to show correlations with the stock
market indicators \cite{Zhang_Fuehres_Gloor_2009}. However these approaches have been
restricted to investor sentiment with only one perspective of macro-economics and are
not complete and flexible in terms explaining complete dynamics that can be extended to
individual stock index for companies. Sprengers et al.
analyzed individual stocks for S\&P 100 companies and tried
correlating tweet features about discussions of the
stock discussions about the particular companies containing the
Ticker symbol \cite{Sprenger}. This paper is an incremental step towards bringing out a
flexible and novel approach combining the search behavior along with the sentiment analysis that is scalable (modified easily) for both individual commodities stocks/ companies. The approach can
be further exploited to make successful hedging strategies making \textit{wisdom of the crowd usable even by a singular investor}.

\vspace{-0.2cm}
In this paper, we present a comprehensive study of relationships over wide range of market securities- commodities such as oil, gold, forex rates of Euro and equity markets such as DJIA and NASDAQ-100 with the dynamic features of the investor behavior as reflected in the opinions emerging on Twitter and trends in the search engine volumes. The summary of the whole study conducted in this paper is provided in the figure \ref{fig:flow_chart}. In section \ref{data_collection} we present data collection and prior processing that explains the terminologies used in the market securities and social mood series. Further in section \ref{results} we present the statistical techniques implemented and discuss the results and draw conclusions. Future prospects of the work are given in section \ref{conclusion}.
\vspace{-0.2cm}
\section {DATA COLLECTION AND PROCESSING}
\label{data_collection}
In this section, we discuss the collection of various financial data series used in this paper.
\vspace{-0.2cm}
\subsection {Tweets Extraction and Processing}

Tweets are made accessible through a simple search of keywords(various market securities in our case) through an application programming interface (API)\footnote {Twitter API is easily accessible at- https://dev.twitter.com/docs. Also Gnip - http://gnip.com/twitter, the premium platform available for purchasing historic and present public firehose of tweets has many investors as financial customers researching in the area, though due to confidentiality issues they are not explicitly named}.
In this work, we have used tweets from period of $15$ months and $10$ days between June 2nd to 13th September 2011. During this period, by querying the Twitter search API for each of the market feature under study say Gold, Euro, Dow etc. we collected $1,964,044$ (by around 0.71M users) English language tweets. Each tweet record contains (a) tweet identifier,(b) date/time of submission(in GMT), (c) language and (d)text. Subsequently the stop words and punctuation are removed and the tweets are grouped for each day (which is the highest time precision window in this study, since we do not group tweets further based on hours/minutes).
\vspace{-0.4cm}
\subsubsection {Tweet Sentiment Extraction}
In order to compute sentiment for any tweet we classify each
incoming tweet everyday into {\it positive} or {\it negative} using naive classifier. For each day total number of positive tweets are aggregated as $Positive_{day}$ and total number of negative tweets as $Negative_{day}$. We have made use of lexicon/ JSON API from Twittersentiment \footnote{https://sites.google.com/site/twittersentimenthelp/},
a service provided by Stanford NLP research group \cite{Alec_Bhayani}. Their training was done over a dataset of 1,600,000 tweets and the classifier achieved an accuracy of about 82.7\%.
Online classifier has made use of Naive Bayesian classification method,
which is one of the successful and highly researched algorithms
for natural language processing classification. It is known to give superior performance to other methods
in context of tweets \cite{Alec_Bhayani}.
Because of the limitations of state of the art natural language processing algorithms, the accuracy of the classifier decreases tremendously when number of moods states (or number of classes) is taken higher. This decrease in sentiment accuracy affects the prediction accuracy as different in the rate of change of sentiment can be measured with significant precision.
Naive Bayesian classification methods have high replicability and few
arbitrary fine tuning elements. Our data has shown that the remaining residual 17.5\% of the tweets (misclassifications) are equally distributed over the two classes, which doesn't affect the overall prediction accuracy as all we are interested in is rate of change of sentiment index over a period of time.

In our dataset roughly 67.14\% of the tweets are positive,
while 32.86\% of the tweets are negative for the market securities under study. This result indicates stock/ commodity discussions to be
much more balanced in terms of agreement than chat and internet board
messages where the ratio of positive to negative from earlier works ranges from 7:1~\cite{Dewally} to 5:1~\cite {Ant_frank}.
Balanced distribution of stock discussion provides us with
more confidence to study information content of the positive and
negative dimensions of discussion about the stock prices on
microblogs.
\vspace{-0.4cm}
\subsubsection {Feature Extraction and  Aggregation}
\label{twitter_terms}
Further positive and negative tweets from each day are aggregated to make weekly time domain indicators which is the time period under study. We selected weekly domain over daily, bi-daily, bi- weekly or monthly as it is the most balanced time resolution to study the effect of investor behavior over model performance accuracy; keeping in-market monetization potential practically impeccable.

\vspace{-0.2cm}
For every week, the value of the security (closing, volatility, volume, weekly returns for each index) is recorded every Friday at closing time of the market trading hours $21:00$ UTC. To explore the relationships between weekly trading and also on the days when market remains closed (weekends, national holidays), we broadly focus on two domains of tweet sentiments- weekday indices and weekend indices (further referred as \textit{WD} and \textit{WK} respectively).
We have carried forward the work of Antweiler et al. for
defining bullishness ($B_t$) for each time domain (time window is \textit{WD} or \textit{WK})
as given by equation \ref{bullishness}:
\vspace{-0.1cm}
\begin{equation}\label{bullishness}
    B_t = \ln{\frac{1+{M_t}^{Positive}}{1+{M_t}^{Negative}}}
\vspace{-0.09cm}
\end{equation}

Where ${M_t}^{Positive}$ and ${M_t}^{Negative}$ represents the number of positive
or negative tweets during a particular time period \textit{WD} or \textit{WK}. Logarithm of bullishness measures
the share of surplus positive signals and also gives more weight to larger number of messages in the specific sentiment group (positive or negative). Message volume is simply defined as natural logarithm of total number of tweets per time domain for a specific security/index. And the agreement among positive and negative tweet messages is defined as:
\vspace{-0.1cm}
\begin {equation}
A_t = 1- \sqrt{(1- \frac{(M_t^{Positive}-
M_t^{Negative})}{(M_t^{Positive}+ M_t^{Negative})}}
\vspace{-0.09cm}
\end {equation}
If $all$ the tweet messages about a particular company are positive (bullish about a company stock) or negative (bearish about a company stock), agreement would be $1$ in that case. Influence of silent tweets days in our study (trading days when no tweeting happens about
particular company) is less than $0.1\%$ which is significantly less
than previous works~\cite{Ant_frank,Sprenger}.
Every market index/ security thus have a total of $10$ potentially causative time series from Twitter: positive \textit{WD}, negative \textit{WD}, bullishness \textit{WD}, message volume \textit{WD}, agreement \textit{WD} and from previous weekend we have positive \textit{WK}, negative \textit{WK}, bullishness \textit{WK}, message volume \textit{WK} and agreement \textit{WK}.

\vspace{-0.2cm}
\subsection{Search Volume Index}
\label{SVI_terms}
To generate search engine lexicon for each of the five securities under study- Oil, DJIA, NASDAQ-100, Gold and Euro; we start by collecting weekly search volumes for specific search terms related to respective sectors like- oil, GLD, Dow-30, nasdaq, oil price etc. as given in Table \ref{tab:search_terms} from \textbf{Google Insights of Search} \footnote{http://www.google.com/insights/search/}. Google provides this open service to access the search volume data at weekly minimum frequency since January $2004$. Next we also take into account the top recommended relevant search terms by Google insights of search, thus expanding the already existing group of search terms.

To further normalize and better understand the computational results, we apply dimension reduction technique of principle component analysis. We are able to reduce the number of variables (uptil $50$ for oil) from search domain by combining similarly behaving time series to create completely uncorrelated co-independent factors- Fact $1$ and Fact $2$.
Principal component analysis (PCA) is a mathematical procedure that uses an orthogonal transformation to convert a set of observations of possibly correlated variables into a set of values of uncorrelated variables called principal components which reveals underlying structure that is responsible for maximum variance. As given in Appendix, Table \ref{tab: Dow_SVI_Factors}, \ref{tab:Euro_SVI_factors}, \ref{tab:Gold_SVI_Factors}, \ref{tab:Nasdaq_SVI_Factors} and \ref{tab:Oil_SVI_Factors} gives the extracted factors by varimax rotation technique to produce orthogonal factors. To identify the factors that cause maximum variance in retweets, we have used Kaiser criterion in which the factors with eigen values greater than 1 are extracted.
\vspace{-0.2cm}
\begin{table}[htbp]
  \centering
  \small
  \caption{Google search Terms for $5$ Securities}
    \begin{tabular}{|c|p{6cm}|}
    \addlinespace
    \toprule
    US Oil Funds & oil commodity, crude oil, oil etfs, curde oil price, oil futures, oil quotes, oil price per barrel, oil prices bloomberg, wti crude oil, oil prices, how much of oil is left, crude oil ticker + 50 more similar terms etc. \\
    \midrule
    DJIA & djia, dow jones industrial average, dow jones, dow, s\&p 500, Stock Market, stock message board \\\hline
    Nasdaq-100 & nasdaq up, djia today, dow futures quote, futures quote, djia quote, nylc, bank of america dividends \\\hline
    Gold   & buy gold, invest in gold US data, invest in gold worldwide, dollar to pound exchange rate, dollar to pound exchange \\\hline
    Euro   & exchange rates converter,dollar euro exchange rate history,rupee exchange rate, oanda currency,rupee exchange,dollar rupee exchange rate,bloomberg live tv,eurusd \\
    \bottomrule
    \end{tabular}%
  \label{tab:search_terms}%
\end{table}%
\vspace{-0.2cm}
\subsection{Financial Market Data}
\label{financial_terms}
We have done analysis in five different sectors oil, DJIA , NASDAQ-$100$, gold and Euro. Most of the data, including all the VIX indices and Euro to USD fedex rates used for analysis are collected from econometrics data from Federal Reserve Bank of St. Louis \footnote{Federal Reserve Economic Data: http://research.stlouisfed.org/fred2/}. Gold prices time series are downloaded from World Gold Council \footnote{http://www.gold.org/investment/statistics/goldpricechart/}. Weekly time series for US oil funds and weekly index movements in DJIA and NASDAQ-$100$ are extracted from Yahoo Finance! API\footnote{http://finance.yahoo.com/}.

The financial features (parameters) available from Yahoo finance under study are opening ($O_t$) and closing
($C_t$) value of the stock/index, highest ($H_t$), lowest ($L_t$) value and volume traded for
the stock/index. In addition returns are defined as difference between the logarithm of closing values of the stock index between the week's Friday and previous week's Friday.
\begin {equation}
 R_t= \{{\ln Close_{(t)}-\ln Close_{(t-1)}}\}\times100
\end {equation}
Trading volume is the logarithm of number of traded shares every week. We estimate
weekly volatility based on intra-day highs and lows using Garman and
Klass volatility measures~\cite{Garman_Klass} given by
the formula:
\begin{equation}
\sigma= \sqrt{\frac{1}{n}\sum{\frac{1}{2}[\ln{\frac{H_t}{L_t}}]^2
- [2\ln{2}-1][\ln{\frac{C_t}{O_t}}]^2}}
\end {equation}

Further in this section we will discuss the various security indices in each of the sector under study.
\subsubsection{Oil}
In this study we have taken USO- United States Oil Fund, an exchange traded fund (ETF) that is one of the highly traded security and strongly tracks movements of light, sweet crude oil purchased and sold at NYSE Arca. We have extracted weekly closing values, volatility and volume parameters from the lexicon. In addition to this we have also taken CBOE OIL volatility index \footnote{http://www.cboe.com/micro/oilvix/introduction.aspx} (further referred as VIX) which is index measure of market's expectation of $30$-day volatility of crude oil prices.
\vspace{-0.1cm}

\subsubsection{DJIA}
 Its an aggregate of $30$ highly traded and influential stock evenly distributed over all sectors. We have taken weekly returns, volatility and volume as parameters under the study. Further we have also extracted CBOE DJIA VIX which is indicative measure of fluctuation in 30-day future index sensitivities.

\vspace{-0.1cm}
\subsubsection{NASDAQ-$100$}
Its an aggregate of the top $100$ stocks from $NASDAQ$ exchange which indexes majority of the technological stocks in the market. For this as well we have taken weekly returns, volatility and volume as the parameters under study. In addition we also extracted CBOE NASDAQ-$100$ VIX which is indicative measure of 30-day ahead index movements.

\vspace{-0.1cm}
\subsubsection{Gold}
We have taken price in US dollar (USD) as its the most traded currency for gold in the world to accurately represent search volumes in each country and related twitter buzz for the precious metal. Further we have extracted Gold ETF VIX as well from CBOE, as indicative of a month ahead fear-gauge in the price of the precious metal.

\vspace{-0.1cm}
\subsubsection{Euro}
We have taken only two parameters- one Euro to USD (US dollar) conversion rates at closing of the market on Friday's eve for every week and other CBOE Euro ETF VIX as measure of 30-day market fear for the same.

\vspace{-0.1cm}
\section{STATISTICAL TECHNIQUES AND RESULTS}
\label{results}
In this section we begin statistical analysis and forecasting performance on each of the financial securities as discussed in the section \ref{financial_terms} from two dynamic components investor behavior comprising of $10$ components from Twitter as discussed in section \ref{twitter_terms} and $1$ or $2$ principle factors from Google SVI as discussed in section \ref{SVI_terms}.
First we identify correlation patterns across various time series at different lagged intervals, further testing the causative relationships of SVI and tweet features on the market securities using econometric technique of Granger's Casuality Analysis. Then we make use of expert model mining system (EMMS) to propose and test the forecasting model and draw performance based conclusions.

\subsection{Correlation and Cross-Correlation Analysis}
We begin our study by identifying pairwise correlation metrics between $10$ Twitter features for each security index given in section \ref{twitter_terms} and the factors derived from SVI search factors as given section \ref{SVI_terms}.

\subsubsection{Technique}
Once we obtain the pearson correlation coefficients, as an evaluation of the lagged response of relationships existing between financial features, Twitter sentiments and the search volumes; we compute cross-correlation at a lag of $\pm$ $7$ week lag to show confidence and effectiveness in results. It also motivates us to look forward in making an accurate forecasting model by picking accurate regressor co-efficient.

For any two series $x = \{x_1,......,x_n\}$ and $ y= \{y_1,......,y_n\}$, the cross correlation lag $\gamma$ at lag k is defined as:

\begin{equation}
\label{eq_cross_ corr}
 \gamma = \frac{\sum_{i} (x_{i+k}- \overline{x})(y_i - \overline{y})}{\sqrt{\sum_{i} (x_{i+k} - \overline{x})^2}\sqrt{\sum_{i} (y_i - \overline{y})^2}}
\end {equation}

In equation \ref{eq_cross_ corr}, $\overline{x}$  $\overline{y}$ are the mean sample values of x and y respectively. Cross-correlation function defined as short for ccf(x,y), is estimate of linear correlation between $ x_{t+k}$ and $y_t$, which means keeping the time series y stationary, we move the time series y backward to forward in time by a lag of k i.e. k= [-7,7] for lags fo 7 weeks in positive and negative direction. Cross-correlation gives the measure of anticipated values of statistically significant relations in a time series \emph{x} which can be made part of the forecasting models discussed ahead.

\subsubsection{Results}
\label{result_corr}
\begin {figure}[h!]
\centering
\includegraphics [scale=0.45] {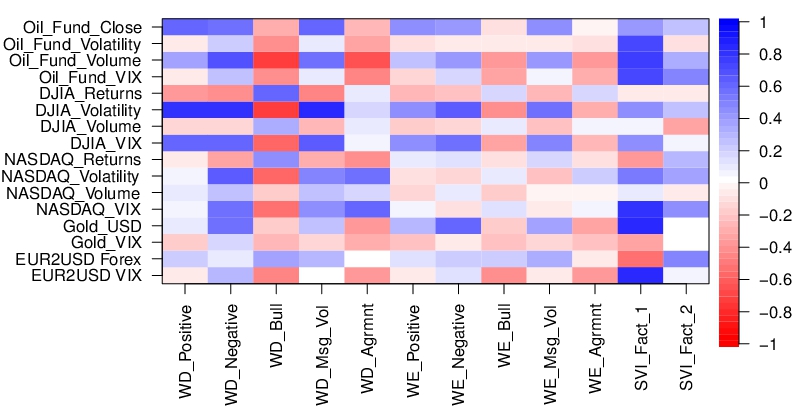}
\caption {Heatmap showing pearson correlation coefficients between security indices vs features from Twitter and SVI factors. (Blue and red correspond to significant correlation values in Figure 3)}
\vspace{-0.1mm}
\label{fig:corr_heatmap}
\end{figure}

The Figure \ref{fig:corr_heatmap} as heatmap along with Figure 3 as radar maps, represents summarized set of pearson correlation results for financial, Twitter and SVI time series after transformation to log scale. Corresponding red and blue sections in both the figures correspond to statistically significant relationships between the dependent variables.

For Twitter features we examine $5$ series- positive, negative, bullishness (Bull), message volume (Msg Vol.) and agreement (Agrmnt) as two cases one as weekday(active market trading) and other as weekend (during market off days). We realize that the overall nature of relationship exhibit varying degree of association. But the clear trend that we observe is that market-off days don't carry high weights when compared to overall data available on comparison to market active days but its still significantly correlated and can be potentially exploited while designing the hedging strategies as discussed later in this paper. Weekday bullishness is one of the important feature out of all others to look out for any investment and show uniformly significant behavior in all the sectors with value of pearson 'r' as high as $- 0.73$ for DJIA's weekly volatility. Another interesting trend that is observed is returns in both DJIA and NASDAQ-$100$ show negative relationship of varying strength with both positive and negative feeds indicating heavy discussion which is more sensitive to message volume on Twitter before fall in the index. But significantly valuable relationship of $0.593$ correlation exists with the returns with our introduced feature term bullishness which is relative measure of positive to negative sentiment of investor community as explained in earlier section \ref{twitter_terms}. NASDAQ doesn't carry any relationship with weekends Twitter discussions on account fast and dispersive behavior of news memes among tech-savvy investors of technological stocks whom are expected to be faster response to news. For volatility indices (VIX) for various securities shows significant negative relation with weekday agreement index which is vector distance between positive and negative discussion about any security as measure of accurate picture of about to happen turbulence/ perceived market risk in the coming weeks; except for DJIA which consists major 30 stocks only which are subjected to highly balanced consistent movements due to heavy trading activity across any time domain.

We find stronger correlation of the principle factors from SVI series uptil 0.826 for commodity funds like for oil, gold and Euro forex rates as compared to index movements of DJIA and NASDAQ-$100$; giving an impression that people tend to search more for commodity funds then stock equities indicating a better understanding of control heuristics of actual market movements from investor behavior. From Figure \ref{fig:corr_heatmap} we can see that the VIX is one of the highly correlated financial feature in all the 5 cases, thus maybe referred as a strong measure of investor behavior though \emph{computational gauge of investor fear}. Another important significant relation that we observed for NASDAQ-$100$ and DJIA is the negative correlation with returns in contrast to positive correlation of volatility, volume and VIX; which is indicative of high search behavior being caused by fall in the index values, increasing more volume in trading making the index movements more volatile.

\vspace{-1mm}

\begin {figure}[]
\centering
\label{fig:ccf_diag}
\subfigure[Oil]{
\includegraphics [scale=0.178] {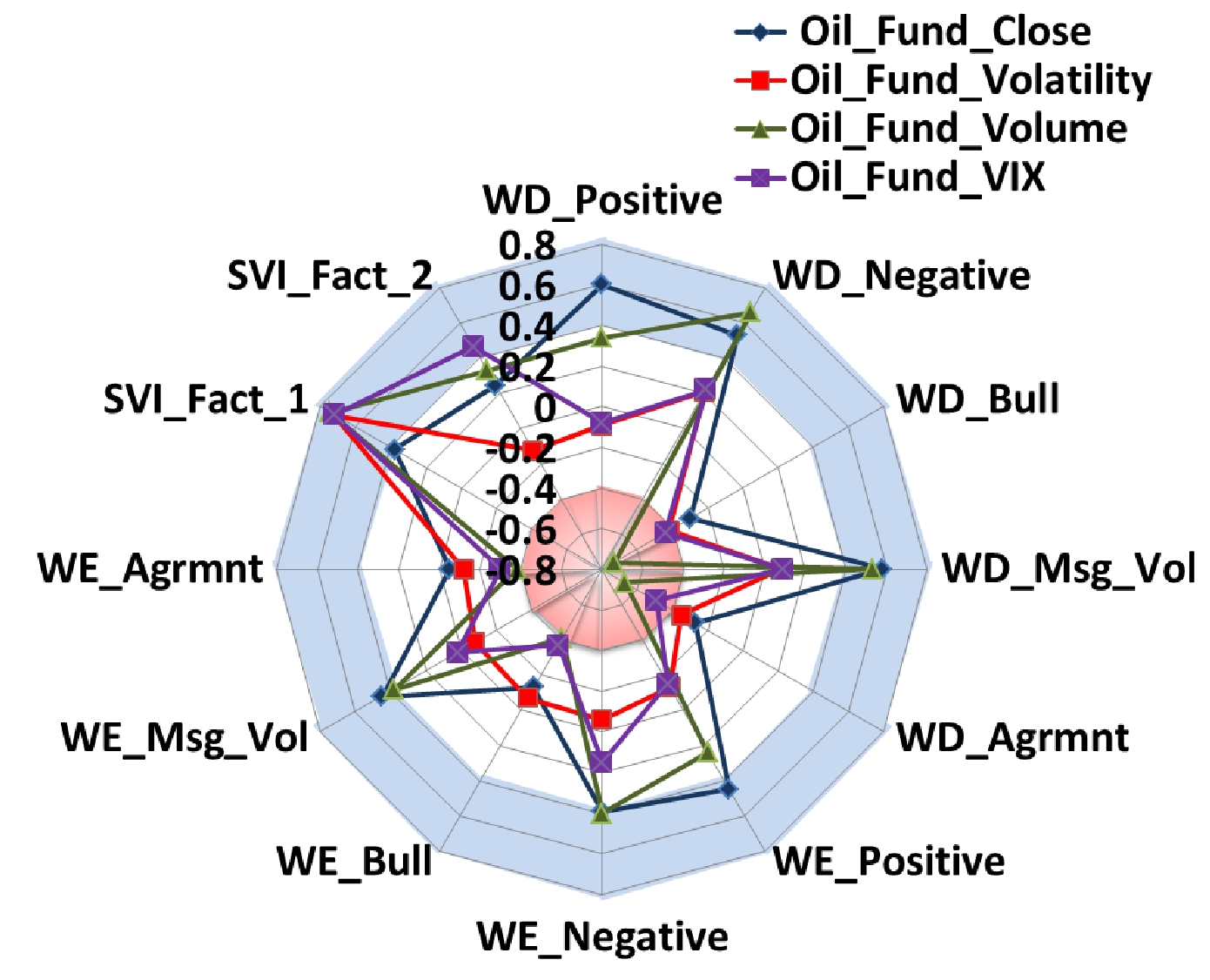}
\label{fig:ccf_oil}
}
\vspace{-2.8mm}
\subfigure[DJIA]{
\includegraphics [scale=0.178] {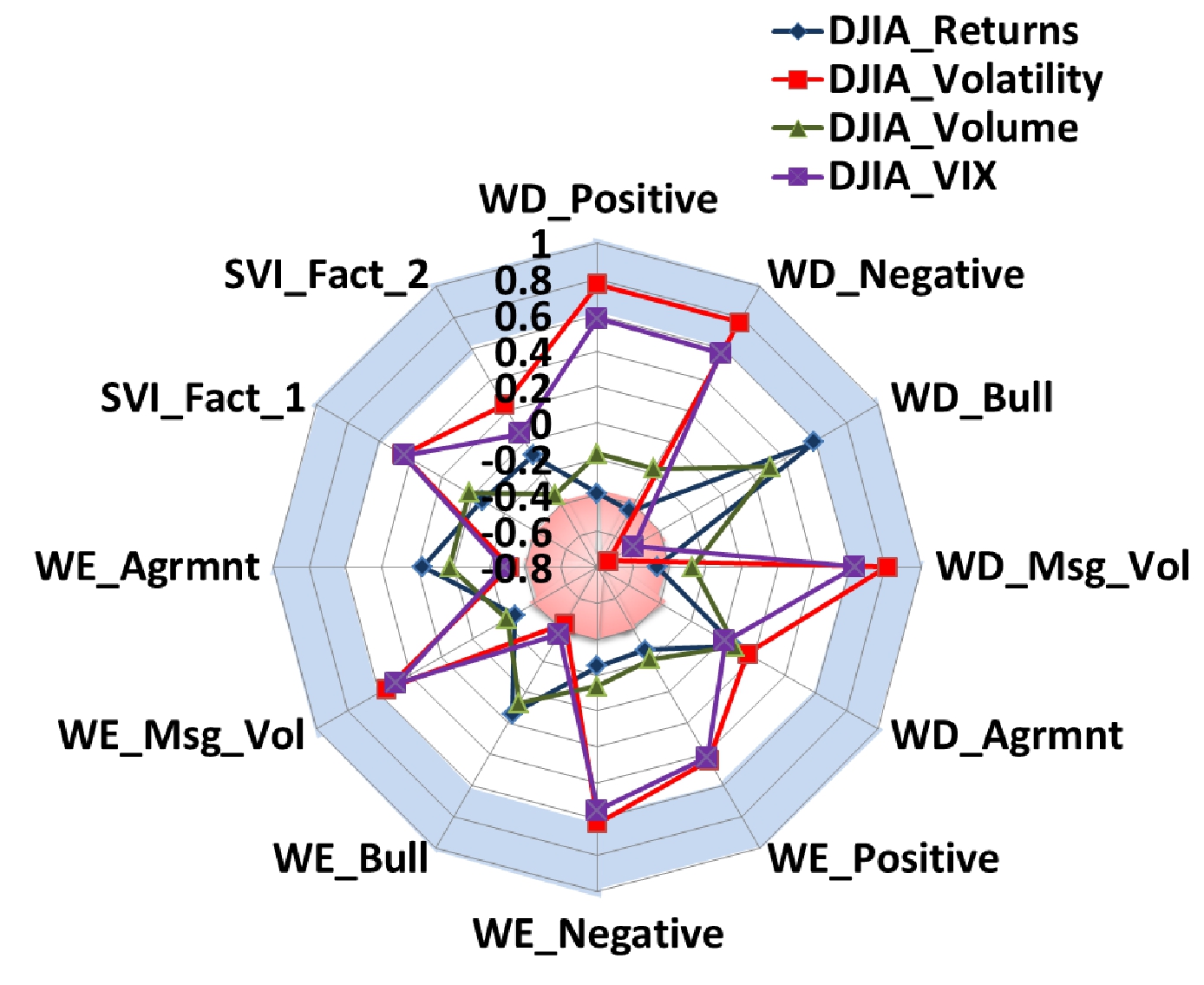}
\label{fig:ccf_gold}
}
\hspace{-3.5mm}
\vspace{-1.5mm}
\subfigure[NASDAQ-100]{
\includegraphics [scale=0.178] {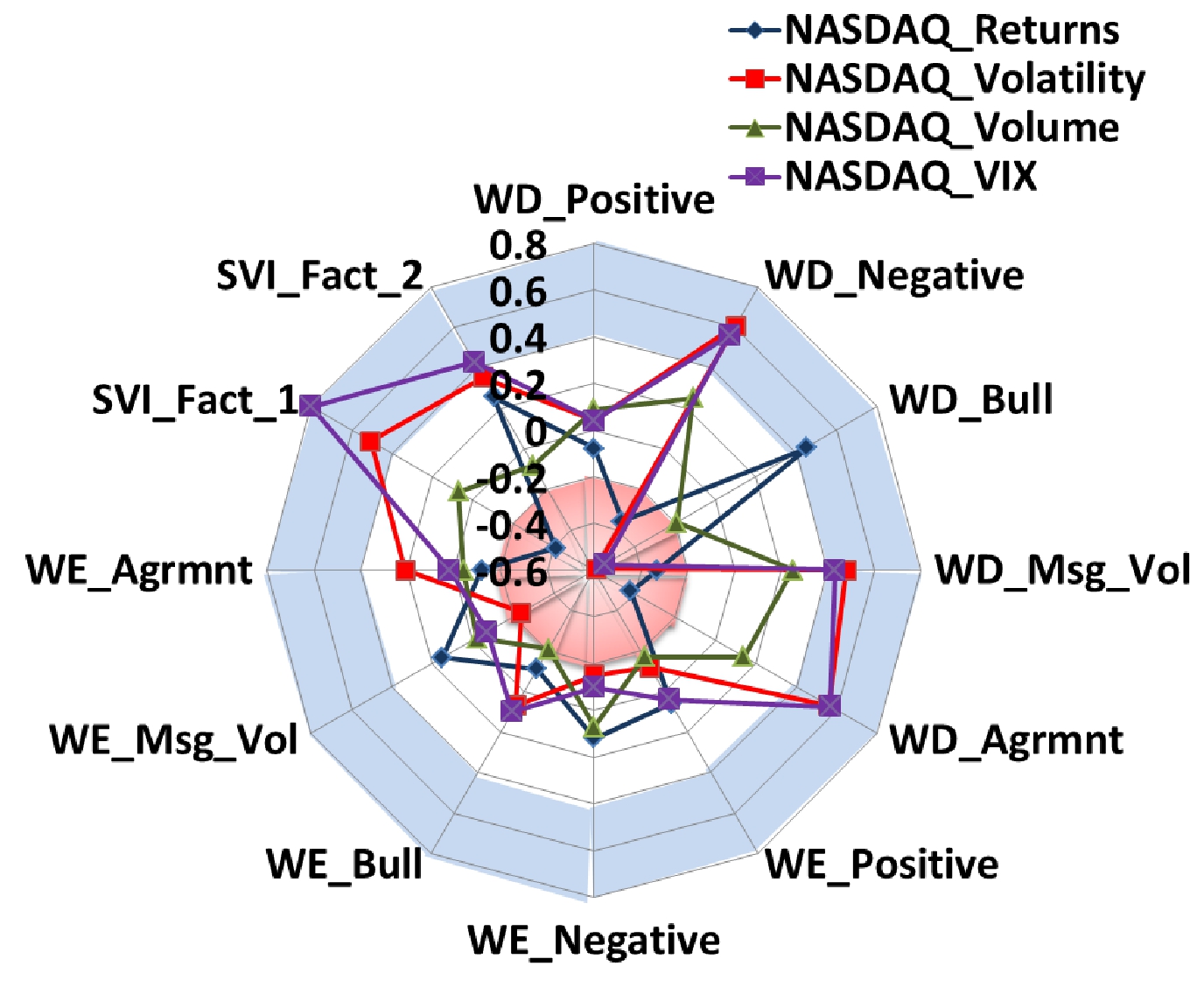}
\label{fig:ccf_djia}
}
\vspace{-0.5mm}
\subfigure[Gold and Euro]{
\includegraphics [scale=0.178] {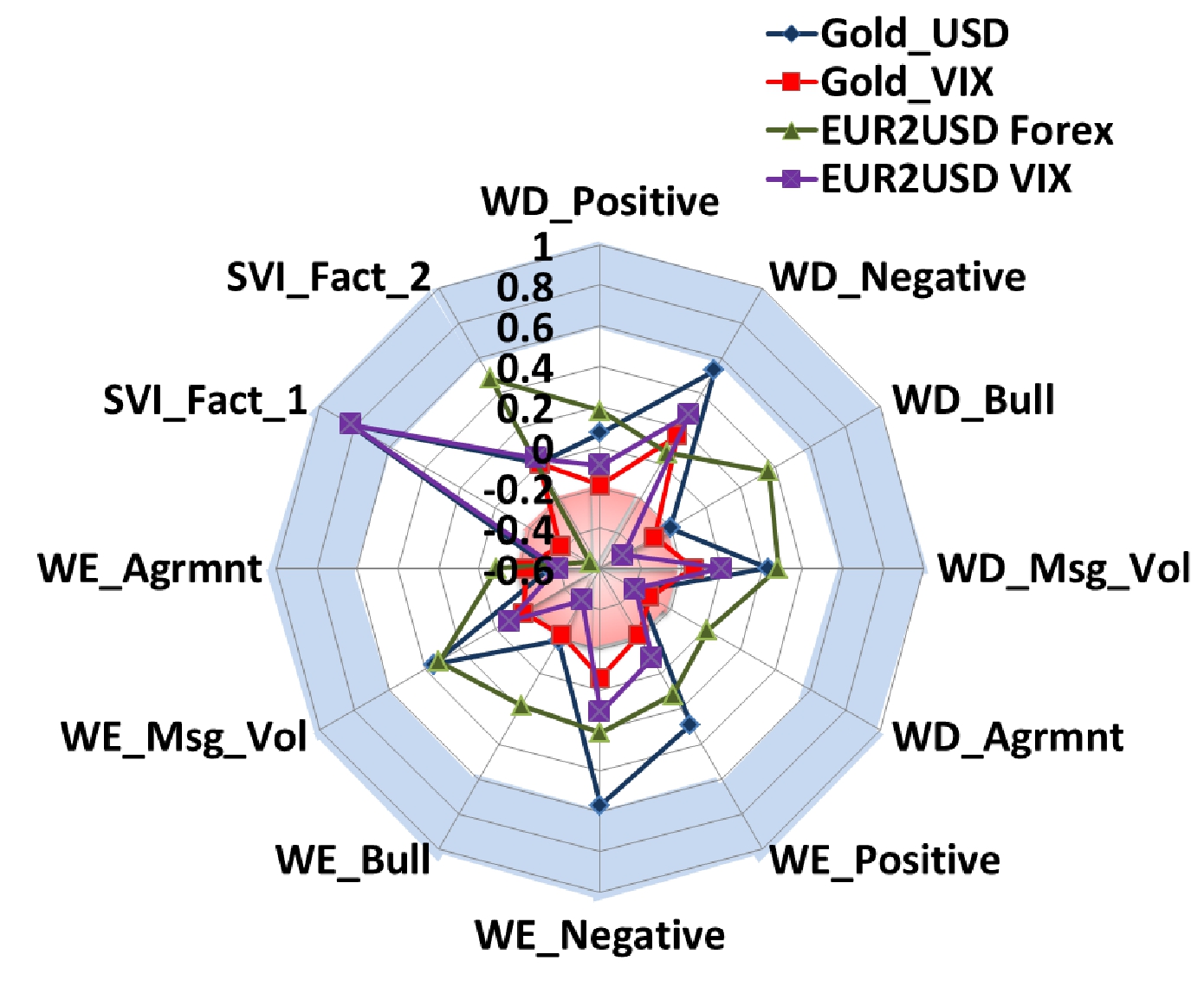}
\label{fig:ccf_nasdaq}
}
\caption[Correlation Radar]{Radar maps showing pearson correlations of Twitter and SVI features vs commodities like oil \subref{fig:ccf_oil}; stock indices like DJIA \subref{fig:ccf_gold} and NASDAQ-100 \subref{fig:ccf_djia}; forex rate of Euro and gold \subref{fig:ccf_nasdaq}. (Blue and red correspond to significant correlation values in Figure ~\ref{fig:corr_heatmap})}
\end{figure}

As we can see in the figure 4 \subref{fig:ccf_oil}, highest correlation is exhibited by oil VIX and SVI which is roughly balanced on both the sides indicating a bi-causative relation in both the directions. Similar observations can be seen for oil fund and oil VIX to Twitter message volume. Tweet message volume have stagnant low slope on the negative lag side which indicates surge in oil related discussion on Twitter consistently prior to actual hike in the price.
For DJIA and NASDAQ-$100$ as observed from figure 4 \subref{fig:ccf_djia} and \subref{fig:ccf_nasdaq}, much balanced correlation factors can be observed for majority of the pairs in both the cases. However, for DJIA significant bend on the negative lag side is observed by volatility in the index for k=-1, indicating a fall of -0.8 correlation in tweet based bullishness atleast a week before the actual market trading. Similar effect is observed for search volumes uptil 4-5 weeks before the actual trading volume increases. NASDAQ-100's correlation activity doesn't give much insights into relationships between the features which maybe due to non- linear associations or significant relations hidden at smaller time domains frequencies as nature of tech-savvy investors of technological stocks. But we can see that bend on positive \emph{k} lag side for volatility with search volumes for a week before and constant increase in bullishness prior \emph{2} weeks before actual surge in volatility. However we leave this area for future exploration.

\begin {figure}[]
\centering
\label{fig:ccf_diag}
\subfigure[Oil]{
\includegraphics [scale=0.195] {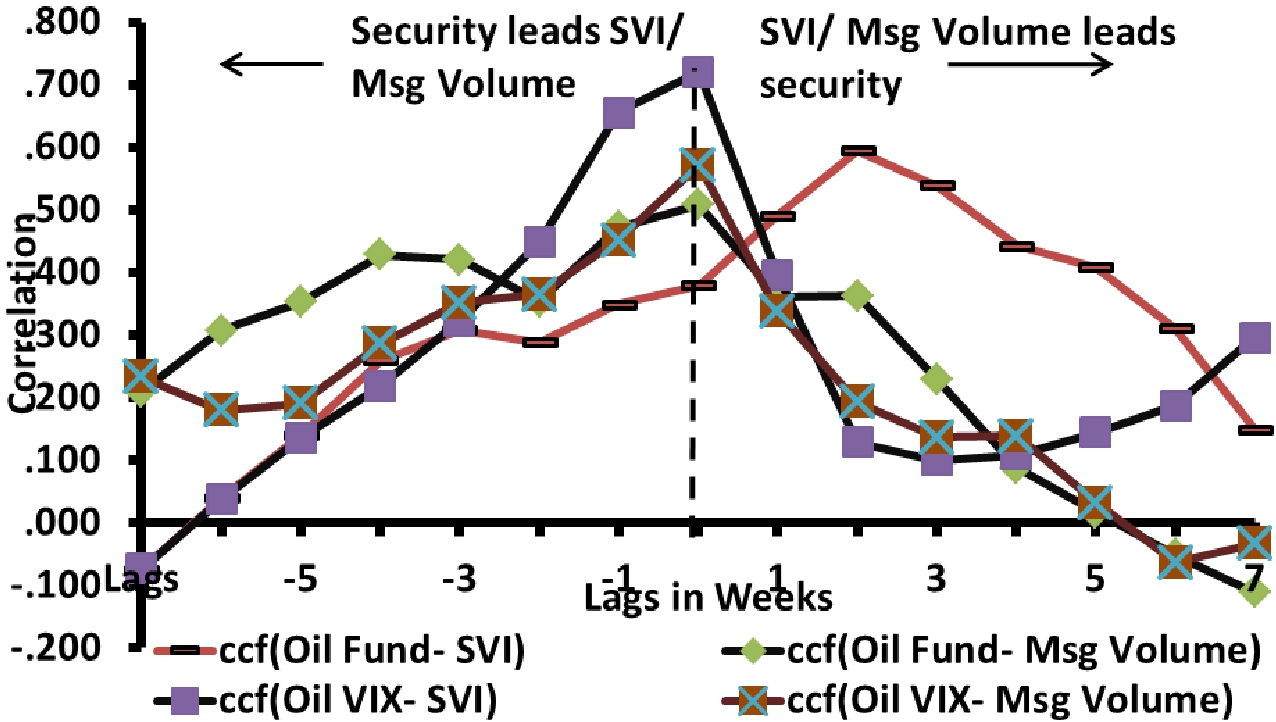}
\label{fig:ccf_oil}
}
\hspace{-3.5mm}
\vspace{-1.5mm}
\subfigure[DJIA]{
\includegraphics [scale=0.195] {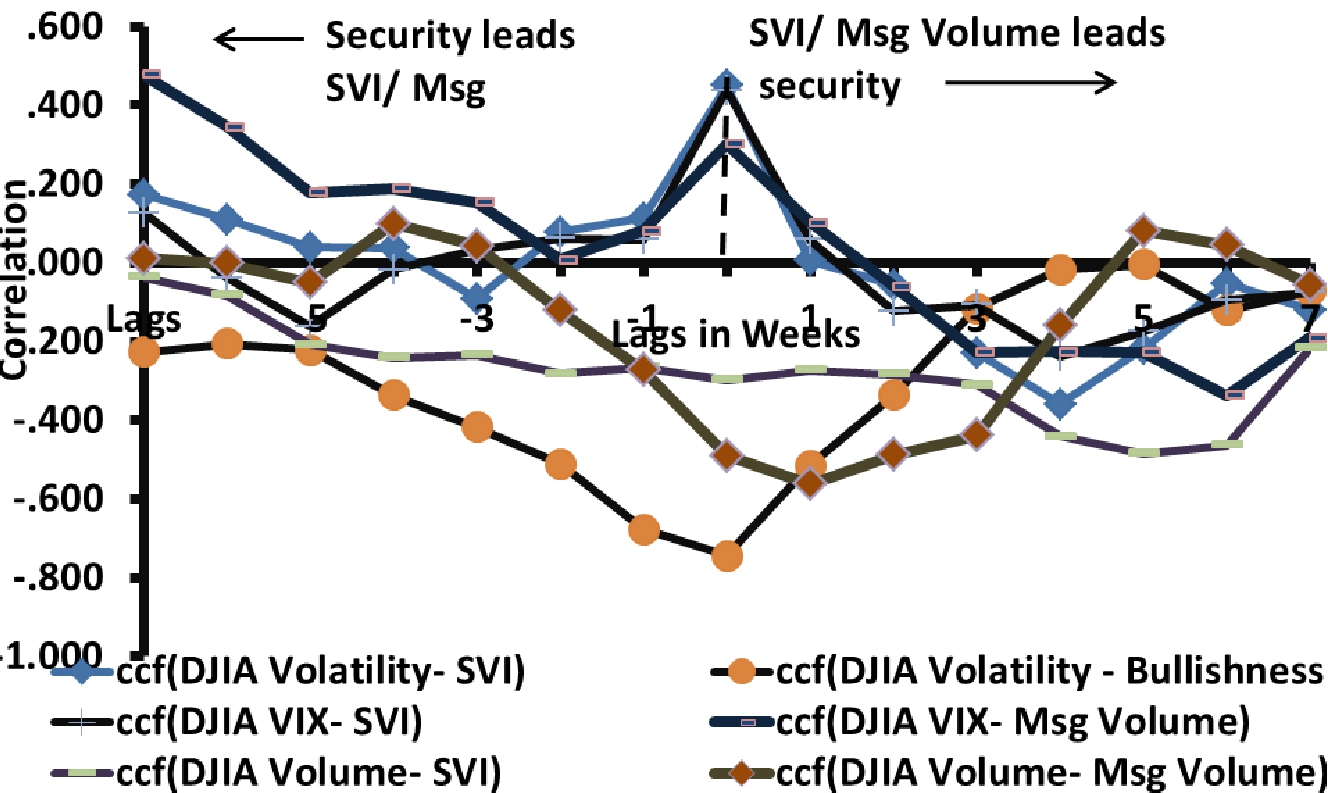}
\label{fig:ccf_djia}
}
\vspace{-2mm}
\subfigure[NASDAQ-100]{
\includegraphics [scale=0.178] {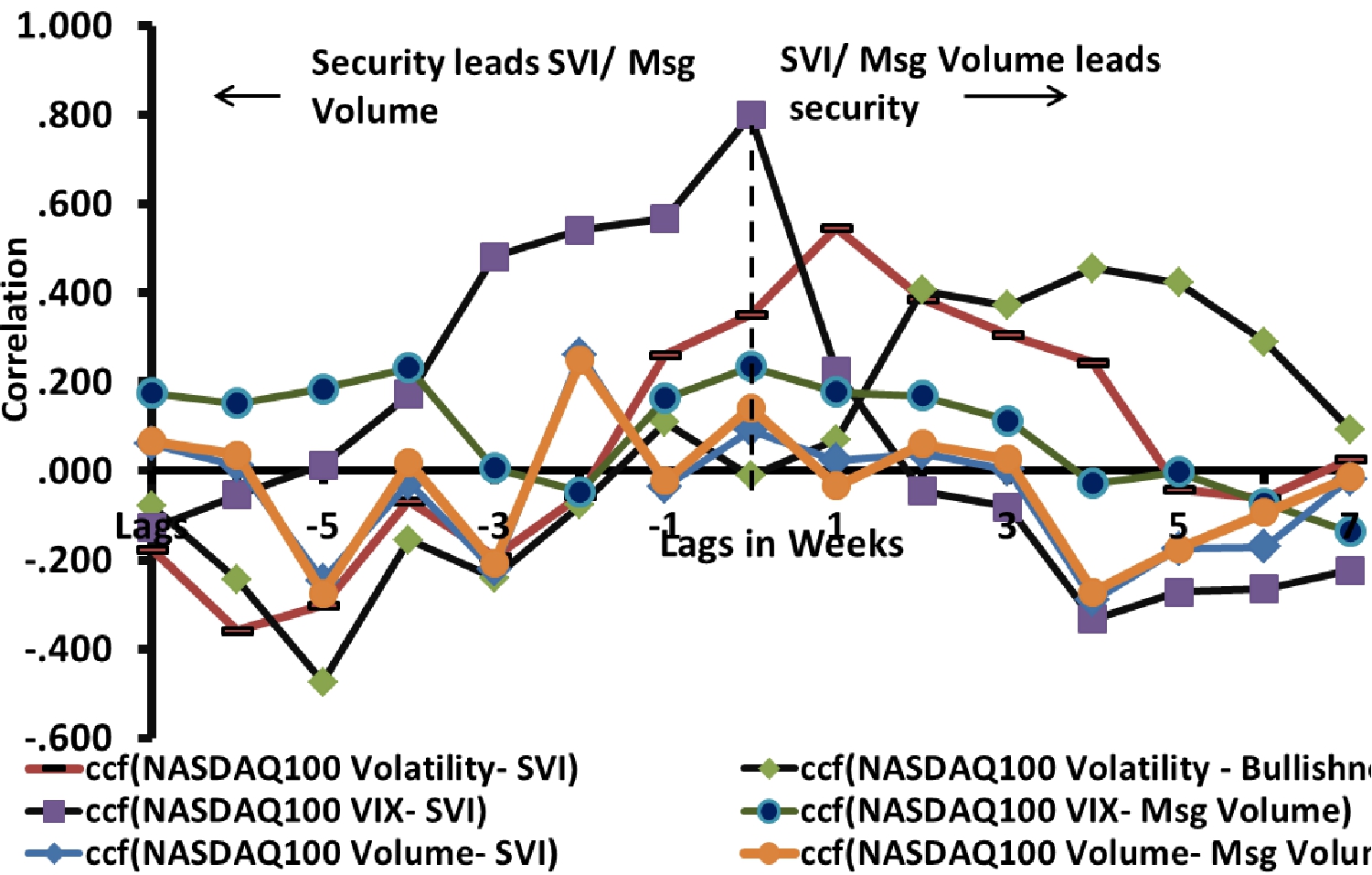}
\label{fig:ccf_nasdaq}
}
\vspace{-2.8mm}
\subfigure[GOLD]{
\includegraphics [scale=0.195] {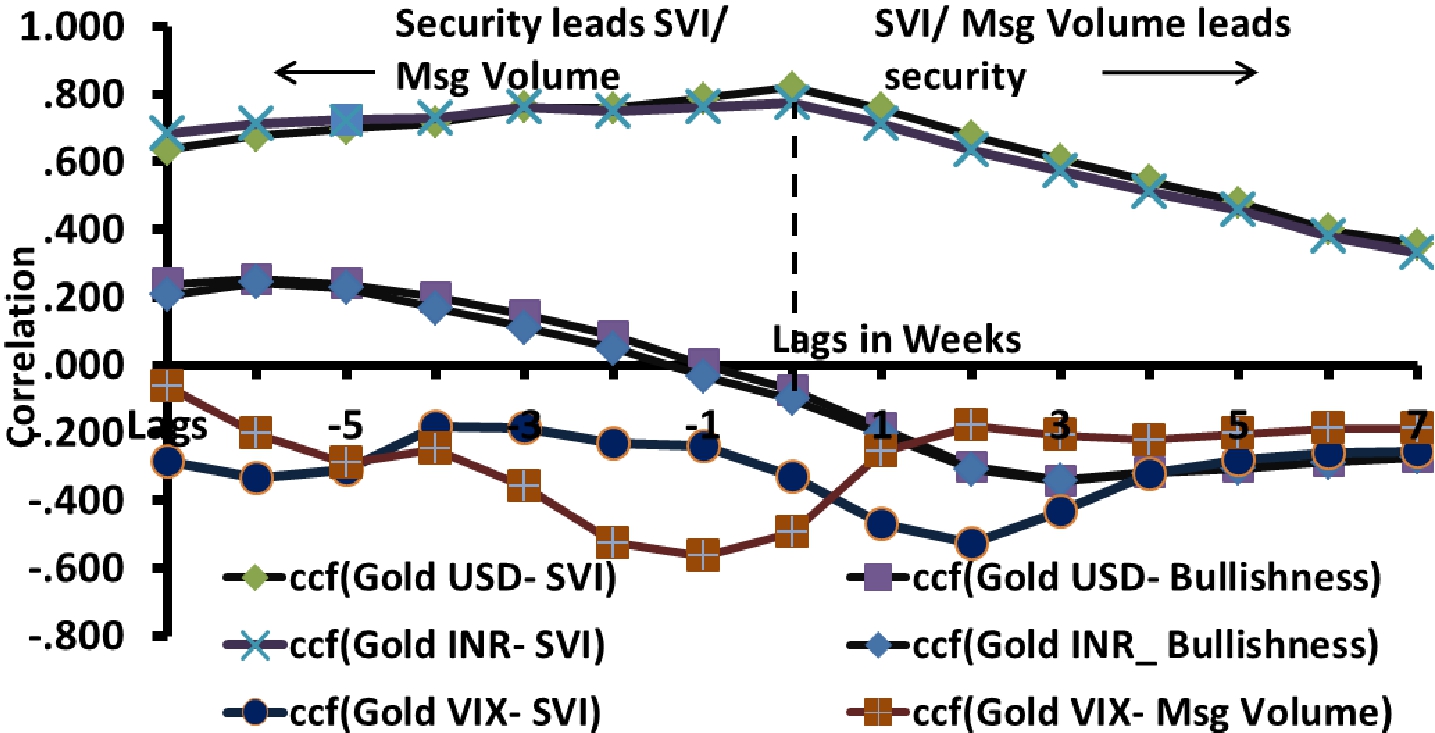}
\label{fig:ccf_gold}
}
\vspace{-1mm}
\subfigure[EURO]{
\includegraphics [scale=0.2] {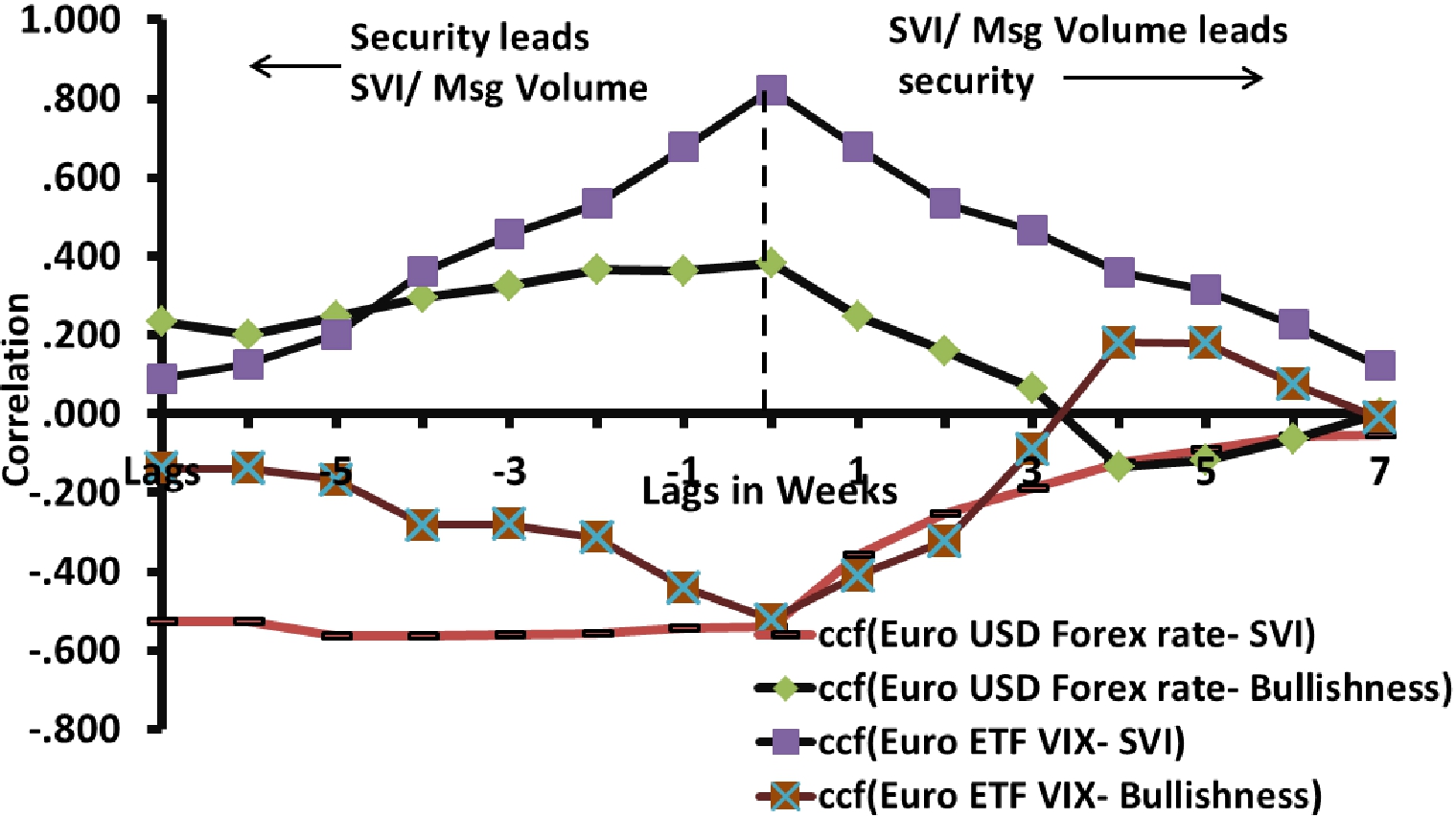}
\label{fig:ccf_euro}
}
\vspace{-1mm}
\caption[Cross Correlation]{Cross Correlation of Twitter and SVI features vs commodities like oil \subref{fig:ccf_oil} and gold \subref{fig:ccf_gold}; stock indices like DJIA \subref{fig:ccf_djia} and NASDAQ-100 \subref{fig:ccf_nasdaq}; and forex rate of Euro \subref{fig:ccf_euro}}
\end{figure}

From figure 4 \subref{fig:ccf_gold} and \subref{fig:ccf_euro} we can see balanced correlation for gold prices and Euro conversion rates. However important conclusions comes when we see behavior of Gold ETF's VIX, which is negative correlation prior one to two weeks; indicating increase gold related tweet discussions before dip in VIX index occurs. But it shows negative correlation at positive lag with search volumes. In contrast we observe a dip in VIX index (fear of buying gold) caused by increased discussion on Twitter as investors consider it as a safe investment, hence the confounding effect further observed in the search volumes.

\vspace{-3.5mm}
\subsection{Granger Causality Analysis}
\label{GCA}
GCA rests on the assumption that if a variable X causes Y then changes in X
will be systematically occur before the changes in Y. We realize
lagged values of X shall bear significant correlation with Y.
However correlation is not necessarily behind causation. Like the earlier approaches by ~\cite{Bollen_Mao_Zeng_2010,Gilbert_Karahalios_2010}
we have made use of GCA to investigate whether one time series is significant in predicting another time
series. GCA is used not to establish statistical causality, but as an economist tool to investigate a statistical pattern of lagged correlation. A similar observation that smoking causes lung cancer is widely accepted; proving it contains carcinogens but itself may not be actual causative of the real event i.e. cancer in this case.
\vspace{-3.5mm}
\subsubsection{Technique}
Let returns $R_t$ be reflective of fast movements in the
stock market. To verify the change in returns with the change in
Twitter features we compare the variance given by following linear
models in equation \ref{eqn:grangers_1} and equation \ref{eqn:grangers_2}.
\vspace{-0.2cm}
\begin{equation}
\label{eqn:grangers_1}
\vspace{-0.1in}
R_t= \alpha +{\Sigma^{n}}_{i=1}\beta_iD_{t-i}+ \epsilon_t
\vspace{-0.1in}
\end{equation}
\vspace{-0.1cm}
\begin{equation}
\label{eqn:grangers_2}
R_t= \alpha+ {\Sigma^{n}}_{i=1} \beta_i D_{t-i}+ {\Sigma^{n}}_{i=1}{\gamma_i X_{i-t}}+ \varepsilon_t
\end{equation}
\vspace{-0.09cm}
Equation \ref{eqn:grangers_1} uses only '$n$' lagged values of $R_t$ , i.e. ($R_{t-1}$,
. . .,$ R_{t-n}$ ) for prediction, while equation \ref{eqn:grangers_2} uses the
$n$ lagged values of both $R_t$ and the tweet features time series
given by $X_{t-1}, . . . , X_{t-n}$. We have taken weekly time window
to validate the casuality performance, hence the lag values
\footnote{{\it lag at k} for any parameter M at $x_{t}$ week is the value
of the parameter prior to $x_{t-k}$ week. For example, value of returns for the month of
April, at the lag of one month will be $return_{april-1}$ which will
be $return_{march}$} will be calculated over the weekly intervals $1,2,...,7$.

\subsubsection{Results}
From the Table ~\ref{tab:grangers}, we can reject the null hypothesis $(H_o)$
that {\it the SVI and Twitter investor behavior do not affect returns in the financial
markets} i.e. $\beta_{1,2,....,n} \neq 0$ with a high level of confidence (high p-values).
However as we see the result applies to only specific negative and
positive tweets (** for p-value $< 0.05$ and * for p-value $< 0.1$
which is 95\% and 99\% confidence interval respectively).
Other features like agreement and message volume do not have
significant casual relationship with the returns of a stock
index (low p-values).

In Table \ref{tab:grangers} we can see that at the lag of one week, almost all the features are significant in predicting changes in the financial features of oil, DJIA, NASDAQ-$100$, gold and Euro. However as we go in the positive lag direction from 1st to 4 weeks, the significance decreases showing Twitter and SVI mood series as Granger's causative of financial features. SVI shows uniform p values i.e. confidence of uptil 99\% for almost all the sectors- both index (DJIA, NASDAQ-$100$) and commodities (gold, oil and forex rate of Euro). Twitter features specially for the indices- DJIA and NASDAQ-$100$ don't significance beyond 2-3 weeks, indicating the dispersive nature of information entropy on the social networks in contrast to the SVI factors.

\vspace{-0.2cm}
\begin{table}[htbp]
  \centering
  \scriptsize
   \addtolength{\tabcolsep}{-2.5pt}
  \caption{Granger's Casuality Analysis- statistical significance (p values) at lags of 1,2,3 and 4 weeks between financial indicators and features of investor behavior (p - value $<$ 0.01:\ddag, p - value $<$ 0.05:\dag, p - value $<$ 0.1:*)}
    \begin{tabular}{|p{0.6cm}|p{0.6cm}|p{0.49cm}|p{0.16cm}p{0.16cm}p{0.16cm}p{0.07cm}|}
    \toprule
    \multicolumn{2}{|p{0.5cm}|}{\multirow{2}[2]{*}{Securities}} & \multicolumn{5}{p{0.5cm}|}{Lag}\\

    \multicolumn{1}{|p{0.5cm}}{} & \multicolumn{1}{p{0.5cm}|}{} & \multicolumn{1}{|p{0.5cm}}{} & \multicolumn{1}{p{0.5cm}}{1} & \multicolumn{1}{p{0.5cm}}{2} & \multicolumn{1}{p{0.5cm}}{3} & \multicolumn{1}{p{0.5cm}|}{4} \\
    \midrule
    \multirow{12}[4]{*}{Oil} & \multirow{6}[2]{*}{Close} & \multicolumn{1}{l|}{Positive} & \multicolumn{1}{l}{\textbf{.009\dag}} & \multicolumn{1}{l}{\textbf{0.1*}} & \multicolumn{1}{l}{0.755} & \multicolumn{1}{l|}{0.238} \\
       &    & \multicolumn{1}{l|}{Negative} & \multicolumn{1}{l}{\textbf{.014\dag}} & \multicolumn{1}{l}{0.352} & \multicolumn{1}{l}{0.666} & \multicolumn{1}{l|}{0.204} \\
       &    & \multicolumn{1}{l|}{Bull} & \multicolumn{1}{l}{\textbf{.014\dag}} & \multicolumn{1}{l}{0.25} & \multicolumn{1}{l}{0.77} & \multicolumn{1}{l|}{0.238} \\
       &    & \multicolumn{1}{l|}{Msg Vol} & \multicolumn{1}{l}{\textbf{.018\dag}} & \multicolumn{1}{l}{\textbf{0.05\dag}} & \multicolumn{1}{l}{0.911} & \multicolumn{1}{l|}{0.397} \\
       &    & \multicolumn{1}{l|}{Agreement} & \multicolumn{1}{l}{\textbf{0.061*}} & \multicolumn{1}{l}{0.521} & \multicolumn{1}{l}{0.89} & \multicolumn{1}{l|}{0.421} \\
       &    & \multicolumn{1}{l|}{SVI} & \multicolumn{1}{l}{\textbf{0.038\dag}} & \multicolumn{1}{l}{0.201} & \multicolumn{1}{l}{\textbf{0.006\dag}} & \multicolumn{1}{l|}{\textbf{0.001\ddag}} \\ \cline{2-7}
       & \multirow{6}[2]{*}{VIX} & \multicolumn{1}{l|}{Positive} & \multicolumn{1}{l}{\textbf{0.048\dag}} & \multicolumn{1}{l}{0.966} & \multicolumn{1}{l}{0.454} & \multicolumn{1}{l|}{0.746} \\
       &    & \multicolumn{1}{l|}{Negative} & \multicolumn{1}{l}{0.6} & \multicolumn{1}{l}{0.683} & \multicolumn{1}{l}{0.303} & \multicolumn{1}{l|}{0.621} \\
       &    & \multicolumn{1}{l|}{Bull} & \multicolumn{1}{l}{\textbf{0.032*}} & \multicolumn{1}{l}{0.819} & \multicolumn{1}{l}{0.364} & \multicolumn{1}{l|}{0.742} \\
       &    & \multicolumn{1}{l|}{Msg Vol} & \multicolumn{1}{l}{\textbf{0.078*}} & \multicolumn{1}{l}{0.701} & \multicolumn{1}{l}{0.706} & \multicolumn{1}{l|}{0.949} \\
       &    & \multicolumn{1}{l|}{Agreement} & \multicolumn{1}{l}{\textbf{0.008\dag}} & \multicolumn{1}{l}{0.804} & \multicolumn{1}{l}{0.411} & \multicolumn{1}{l|}{0.957} \\
       &    & SVI & \textbf{0.00002\ddag} & \textbf{0.001\ddag} & \textbf{0.037\dag} & \textbf{0.07*} \\ \hline \hline
    \multirow{12}[4]{*}{DJIA} & \multirow{6}[2]{*}{Return} & \multicolumn{1}{l|}{Positive} & \multicolumn{1}{l}{0.675} & \multicolumn{1}{l}{0.601} & \multicolumn{1}{l}{0.986} & \multicolumn{1}{l|}{0.266} \\
       &    & \multicolumn{1}{l|}{Negative} & \multicolumn{1}{l}{\textbf{0.065*}} & \multicolumn{1}{l}{\textbf{0.056*}} & \multicolumn{1}{l}{0.996} & \multicolumn{1}{l|}{0.331} \\
       &    & \multicolumn{1}{l|}{Bull} & \multicolumn{1}{l}{0.38} & \multicolumn{1}{l}{0.442} & \multicolumn{1}{l}{0.991} & \multicolumn{1}{l|}{0.305} \\
       &    & \multicolumn{1}{l|}{Msg Vol} & \multicolumn{1}{l}{\textbf{0.052*}} & \multicolumn{1}{l}{0.608} & \multicolumn{1}{l}{0.947} & \multicolumn{1}{l|}{0.237} \\
       &    & \multicolumn{1}{l|}{Agreement} & \multicolumn{1}{l}{0.264} & \multicolumn{1}{l}{0.243} & \multicolumn{1}{l}{0.826} & \multicolumn{1}{l|}{0.552} \\
       &    & \multicolumn{1}{l|}{SVI} & \multicolumn{1}{l}{\textbf{0.021\dag}} & \multicolumn{1}{l}{\textbf{0.053*}} & \multicolumn{1}{l}{\textbf{0.021\dag}} & \multicolumn{1}{l|}{\textbf{0.057*}} \\ \cline{2-7}
       & \multirow{6}[2]{*}{VIX} & \multicolumn{1}{l|}{Positive} & \multicolumn{1}{l}{0.461} & \multicolumn{1}{l}{0.501} & \multicolumn{1}{l}{0.936} & \multicolumn{1}{l|}{0.683} \\
       &    & \multicolumn{1}{l|}{Negative} & \multicolumn{1}{l}{\textbf{0.024*}} & \multicolumn{1}{l}{0.286} & \multicolumn{1}{l}{0.91} & \multicolumn{1}{l|}{0.388} \\
       &    & \multicolumn{1}{l|}{Bull} & \multicolumn{1}{l}{0.38} & \multicolumn{1}{l}{0.527} & \multicolumn{1}{l}{0.672} & \multicolumn{1}{l|}{0.583} \\
       &    & \multicolumn{1}{l|}{Msg Vol} & \multicolumn{1}{l}{\textbf{0.033*}} & \multicolumn{1}{l}{\textbf{0.05*}} & \multicolumn{1}{l}{0.666} & \multicolumn{1}{l|}{0.97} \\
       &    & \multicolumn{1}{l|}{Agreement} & \multicolumn{1}{l}{0.427} & \multicolumn{1}{l}{0.436} & \multicolumn{1}{l}{0.616} & \multicolumn{1}{l|}{0.752} \\
       &    & \multicolumn{1}{l|}{SVI} & \multicolumn{1}{l}{\textbf{0.015\dag}} & \multicolumn{1}{l}{\textbf{0.06*}} & \multicolumn{1}{l}{\textbf{0.017\dag}} & \multicolumn{1}{l|}{\textbf{0.03\dag}} \\\hline \hline
    \multirow{12}[4]{*}{Nasdaq} & \multirow{6}[2]{*}{Return} & \multicolumn{1}{l|}{Positive} & \multicolumn{1}{l}{\textbf{0.088*}} & \multicolumn{1}{l}{\textbf{0.017\dag}} & \multicolumn{1}{l}{\textbf{0.049\dag}} & \multicolumn{1}{l|}{\textbf{0.1*}} \\
       &    & \multicolumn{1}{l|}{Negative} & \multicolumn{1}{l}{0.737} & \multicolumn{1}{l}{\textbf{0.017\dag}} & \multicolumn{1}{l}{\textbf{0.076*}} & \multicolumn{1}{l|}{\textbf{0.064*}} \\
       &    & \multicolumn{1}{l|}{Bull} & \multicolumn{1}{l}{\textbf{0.061*}} & \multicolumn{1}{l}{\textbf{0.024\dag}} & \multicolumn{1}{l}{0.213} & \multicolumn{1}{l|}{0.136} \\
       &    & \multicolumn{1}{l|}{Msg Vol} & \multicolumn{1}{l}{0.253} & \multicolumn{1}{l}{0.218} & \multicolumn{1}{l}{0.043} & \multicolumn{1}{l|}{0.473} \\
       &    & \multicolumn{1}{l|}{Agreement} & \multicolumn{1}{l}{\textbf{0.091*}} & \multicolumn{1}{l}{0.31} & \multicolumn{1}{l}{0.988} & \multicolumn{1}{l|}{0.245} \\
       &    & \multicolumn{1}{l|}{SVI} & \multicolumn{1}{l}{\textbf{0.091*}} & \multicolumn{1}{l}{\textbf{0.081*}} & \multicolumn{1}{l}{\textbf{0.064*}} & \multicolumn{1}{l|}{\textbf{0.091*}} \\\cline{2-7}
       & \multirow{6}[2]{*}{VIX} & \multicolumn{1}{l|}{Positive} & \multicolumn{1}{l}{\textbf{0.076*}} & \multicolumn{1}{l}{\textbf{0.086*}} & \multicolumn{1}{l}{\textbf{0.025\dag}} & \multicolumn{1}{l|}{\textbf{0.042\dag}} \\
       &    & \multicolumn{1}{l|}{Negative} & \multicolumn{1}{l}{\textbf{0.001\ddag}} & \multicolumn{1}{l}{0.31} & \multicolumn{1}{l}{0.893} & \multicolumn{1}{l|}{0.128} \\
       &    & \multicolumn{1}{l|}{Bull} & \multicolumn{1}{l}{\textbf{0.043\dag}} & \multicolumn{1}{l}{0.241} & \multicolumn{1}{l}{\textbf{0.021\dag}} & \multicolumn{1}{l|}{\textbf{0.04\dag}} \\
       &    & \multicolumn{1}{l|}{Msg Vol} & \multicolumn{1}{l}{\textbf{0.179}} & \multicolumn{1}{l}{0.427} & \multicolumn{1}{l}{\textbf{0.024\dag}} & \multicolumn{1}{l|}{0.148} \\
       &    & \multicolumn{1}{l|}{Agreement} & \multicolumn{1}{l}{\textbf{0.019\dag}} & \multicolumn{1}{l}{0.229} & \multicolumn{1}{l}{0.278} & \multicolumn{1}{l|}{\textbf{0.093*}} \\
       &    & \multicolumn{1}{l|}{SVI} & \multicolumn{1}{l}{\textbf{0.0002\ddag}} & \multicolumn{1}{l}{\textbf{0.002\ddag}} & \multicolumn{1}{l}{\textbf{0.02\dag}} & \multicolumn{1}{l|}{\textbf{0.054*}} \\ \hline \hline
    \multirow{12}[4]{*}{Gold} & \multirow{6}[2]{*}{Price USD} & \multicolumn{1}{l|}{Positive} & \multicolumn{1}{l}{0.136} & \multicolumn{1}{l}{0.331} & \multicolumn{1}{l}{0.631} & \multicolumn{1}{l|}{0.41} \\
       &    & \multicolumn{1}{l|}{Negative} & \multicolumn{1}{l}{0.56} & \multicolumn{1}{l}{0.712} & \multicolumn{1}{l}{0.807} & \multicolumn{1}{l|}{0.66} \\
       &    & \multicolumn{1}{l|}{Bull} & \multicolumn{1}{l}{\textbf{0.004\ddag}} & \multicolumn{1}{l}{\textbf{0.023\dag}} & \multicolumn{1}{l}{\textbf{0.028\dag}} & \multicolumn{1}{l|}{\textbf{0.058*}} \\
       &    & \multicolumn{1}{l|}{Msg Vol} & \multicolumn{1}{l}{\textbf{0.027\dag}} & \multicolumn{1}{l}{\textbf{0.1*}} & \multicolumn{1}{l}{0.625} & \multicolumn{1}{l|}{0.557} \\
       &    & \multicolumn{1}{l|}{Agreement} & \multicolumn{1}{l}{\textbf{0.035\dag}} & \multicolumn{1}{l}{\textbf{0.009\ddag}} & \multicolumn{1}{l}{\textbf{0.015\dag}} & \multicolumn{1}{l|}{\textbf{0.045\dag}} \\
       &    & \multicolumn{1}{l|}{SVI} & \multicolumn{1}{l}{\textbf{0.0001\ddag}} & \multicolumn{1}{l}{\textbf{0.00034\ddag}} & \multicolumn{1}{l}{\textbf{0.00041\ddag}} & \multicolumn{1}{l|}{\textbf{0.001\ddag}} \\\cline{2-7}
       & \multirow{6}[2]{*}{VIX} & \multicolumn{1}{l|}{Positive} & \multicolumn{1}{l}{\textbf{0.083*}} & \multicolumn{1}{l}{\textbf{0.11*}} & \multicolumn{1}{l}{0.192} & \multicolumn{1}{l|}{\textbf{0.05\dag}} \\
       &    & \multicolumn{1}{l|}{Negative} & \multicolumn{1}{l}{\textbf{0.1*}} & \multicolumn{1}{l}{0.454} & \multicolumn{1}{l}{\textbf{0.1*}} & \multicolumn{1}{l|}{\textbf{0.033\dag}} \\
       &    & \multicolumn{1}{l|}{Bull|} & \multicolumn{1}{l}{0.385} & \multicolumn{1}{l}{0.641} & \multicolumn{1}{l}{0.509} & \multicolumn{1}{l|}{0.755} \\
       &    & \multicolumn{1}{l|}{Msg Vol} & \multicolumn{1}{l}{0.793} & \multicolumn{1}{l}{\textbf{0.1*}} & \multicolumn{1}{l}{0.305} & \multicolumn{1}{l|}{0.256} \\
       &    & \multicolumn{1}{l|}{Agreement} & \multicolumn{1}{l}{\textbf{0.1*}} & \multicolumn{1}{l}{0.385} & \multicolumn{1}{l}{0.493} & \multicolumn{1}{l|}{0.184} \\
       &    & \multicolumn{1}{l|}{SVI} & \multicolumn{1}{l}{0.414} & \multicolumn{1}{l}{\textbf{0.059*}} & \multicolumn{1}{l}{\textbf{0.057*}} & \multicolumn{1}{l|}{\textbf{0.05\dag}} \\\hline \hline
    \multirow{12}[4]{*}{Euro} & \multirow{6}[2]{*}{EURUSD} & Positive & \textbf{0.051*} & \textbf{0.11*} & \textbf{0.1*} & 0.336 \\
       &    & Negative & \textbf{0.043*} & 0.51 & 0.249 & 0.561 \\
       &    & Bull & \textbf{0.069*} & 0.754 & 0.521 & 0.497 \\
       &    & Msg Vol & \textbf{0.1*} & 0.439 & \textbf{0.1*} & 0.157 \\
       &    & Agreement & 0.944 & 0.985 & 0.62 & 0.399 \\
       &    & SVI & \textbf{0.00001\ddag} & \textbf{0.00006\ddag} & \textbf{0.00008\ddag} & \textbf{0.0001\ddag} \\\cline{2-7}
       & \multirow{6}[2]{*}{VIX} & Positive & \textbf{0.1*} & \textbf{0.085*} & \textbf{0.092*} & 0.431 \\
       &    & Negative & \textbf{0.028\dag} & \textbf{0.011\dag} & \textbf{0.034\dag} & 0.068 \\
       &    & Bull & 0.498 & \textbf{0.1*} & \textbf{0.1*} & 0.797 \\
       &    & Msg Vol & 0.443 & 0.256 & 0.987 & 0.213 \\
       &    & Agreement & 0.384 & 0.587 & 0.55 & 0.557 \\
       &    & SVI & \textbf{0.091*} & \textbf{0.0001\ddag} & \textbf{0.002\ddag} & \textbf{0.003\ddag} \\\hline
    \bottomrule
    \end{tabular}%
  \label{tab:grangers}%
\end{table}%

\vspace{-1mm}
\subsection{EMMS model for Forecasting Analysis of Financial features}
\label{emms}
In this section we work upon the perennial question of \textit{how much? and how good?} are these features proposed in the earlier sections can be useful to make accurate forecasts of financial indicators. For the same purpose we have used Expert Model Mining System (EMMS) which incorporates a set
of competing methods such as Exponential Smoothing (ES), Auto Regressive
Integrated Moving Average (ARIMA) and seasonal ARIMA models. These
methods are widely used in financial modeling to predict the values of
stocks/ bonds/ commodities/etc \cite{Pegels,Box}. These methods are
suitable for constant level, additive trend or multiplicative trend and
with either no seasonality, additive seasonality, or multiplicative
seasonality.

\subsubsection{Technique}

Selection criterion for the EMMS is MAPE and stationary \emph{R squared} which is measure of how good is the model under consideration as comapred to the baseline model \cite{Harvey_89}. The stationary R-squared can be negative with range $(-\infty, 1]$. A negative R-squared value means that the model under consideration is worse than the baseline model. Zero R-squared means that the model under consideration is as good or bad as the baseline model. Positive R-squared means that the model under consideration is better than the baseline model. Mean
absolute percentage error (MAPE) is mean residuals (difference between fit
value and observed value in percentage). To show the performance of
tweet features in prediction model, we have applied the EMMS twice - first
with
SVI and Twitter sentiment features as independent predictor events and second time without
them. This provides us with a quantitative comparison of improvement in
the prediction using tweet features.

ARIMA (p,d,q) in theory and practice, are the most general class of
models for forecasting a time series data, which is subsequently
stationarized by series of transformation such as differencing or
logging of the series $Y_i$. For a non-seasonal ARIMA (p,d,q) model- p is
autoregressive term, d is number of non-seasonal differences and q is
the number of lagged forecast errors in the predictive equation.  A
stationary time series $\Delta Y$ differences d times has stochastic
component:
\vspace{-0.2cm}
\begin{equation}
    \Delta Y_i \hookrightarrow Normal(\mu_i,\sigma^2)
\end{equation}
\vspace{-0.01cm}
Where $\mu_i$ and $\sigma^2$   are the mean and variance of normal
distribution, respectively. The systematic component is modeled as:
\vspace{-0.09cm}
\begin{equation}
\vspace{-0.1in}
    \begin {split}
    \mu_i= \alpha_i\Delta Y_{i-1}+.....+ \alpha_p \Delta Y_{i-p}
    + \theta_i\varepsilon_{i-1}+.....\\+ \theta_i
    \varepsilon_{i-q}
    \end {split}
\end{equation}
\vspace{-0.01cm}
Where, $\Delta Y$ the lag-p observations from the stationary time series
with associated parameter vector $\alpha$ and $\epsilon_i$ the lagged
errors of order q, with associated parameter vector. The expected value
is the mean of simulations from the stochastic component,
\vspace{-0.09cm}
\begin{equation}
\begin {split} E(Y_(i )=\mu_i= \alpha_i\Delta
Y_{i-1}+.....+ \alpha_p\Lambda Y_{i-p}+ \theta_i\varepsilon_{i-1}
\\ +.....+ \theta_i \varepsilon_{i-q} (2) \end {split}
\end{equation}
\vspace{-0.01cm}
Seasonal ARIMA model is of form ARIMA (p ,d ,q) (P,D,Q) where P
specifies the seasonal autoregressive order, D is the seasonal differencing order
and Q is the moving average order. Another advantage of EMMS model is
that it is a stepwise forecasting process which automatically selects the most significant predictors among all
other Twitter sentiment series and SVI features.

\subsubsection{Results}
Model equation for two cases are given below as equation \ref{eqn:Without Predictors} for forecasting without predictors and equation \ref{eqn:With Predictors} for forecasting with predictors. In these equations Y is the financial feature- oil, gold, DJIA etc. and X represents the investor mood series from the SVI and Twitter features.
\vspace{-0.09cm}
\begin{equation}
\label{eqn:Without Predictors}
Without Predictors: Y_t= \alpha + {{\Sigma^{n}}_{i=1}} \beta_i Y_{t=i} + \epsilon_t
\vspace{-0.2cm}
\end{equation}
\vspace{-0.9cm}

\begin{equation}
\vspace{-0.1cm}
\label{eqn:With Predictors}
With Predictors: Y_t= \alpha + {{\Sigma^{n}}_{i=1}} \beta_i Y_{t=i}+ {{\Sigma^{n}}_{i=1}} \gamma_i X_{t=i} + \epsilon_t
\vspace{-0.2cm}
\end{equation}

\begin{table}[htbp]
  \centering
  \scriptsize
 \caption{Forecasting results for the financial securities}
    \begin{tabular}{|c|r|r|r|r|}
    \toprule
    \multicolumn{2}{|c|}{Market Securities} & \multicolumn{1}{c|}{Predictors} & \multicolumn{1}{c|}{MAPE} & \multicolumn{1}{c|}{Direction} \\ \hline
    \midrule
    \multirow{5}[6]{*}{US Oil Funds} &    &    &    &  \\
       & \multicolumn{1}{c|}{\multirow{2}[2]{*}{Index}} & \multicolumn{1}{c|}{Yes} & \multicolumn{1}{c|}{2.3202} & \multicolumn{1}{c|}{75} \\
       & \multicolumn{1}{c|}{} & \multicolumn{1}{c|}{No} & \multicolumn{1}{c|}{2.4203} & \multicolumn{1}{c|}{62.5} \\ \cline{2-5}
       & \multicolumn{1}{c|}{\multirow{2}[2]{*}{VIX}} & \multicolumn{1}{c|}{Yes} & \multicolumn{1}{c|}{4.5592} & \multicolumn{1}{c|}{75} \\
       & \multicolumn{1}{c|}{} & \multicolumn{1}{c|}{No} & \multicolumn{1}{c|}{5.1218} & \multicolumn{1}{c|}{56.3} \\ \hline\hline
    \multirow{4}[4]{*}{DJIA} & \multicolumn{1}{c|}{\multirow{2}[2]{*}{Index}} & \multicolumn{1}{c|}{Yes} & \multicolumn{1}{c|}{0.8557} & \multicolumn{1}{c|}{94.3} \\
       & \multicolumn{1}{c|}{} & \multicolumn{1}{c|}{No} & \multicolumn{1}{c|}{1.1698} & \multicolumn{1}{c|}{60} \\ \cline{2-5}
       & \multicolumn{1}{c|}{\multirow{2}[2]{*}{VIX}} & \multicolumn{1}{c|}{Yes} & \multicolumn{1}{c|}{5.3017} & \multicolumn{1}{c|}{82.9} \\
       & \multicolumn{1}{c|}{} & \multicolumn{1}{c|}{No} & \multicolumn{1}{c|}{5.6943} & \multicolumn{1}{c|}{62.9} \\ \hline\hline
    \multirow{4}[4]{*}{NASDAQ-100} & \multicolumn{1}{c|}{\multirow{2}[2]{*}{Index}} & \multicolumn{1}{c|}{Yes} & \multicolumn{1}{c|}{1.3235} & \multicolumn{1}{c|}{90} \\
       & \multicolumn{1}{c|}{} & \multicolumn{1}{c|}{No} & \multicolumn{1}{c|}{1.3585} & \multicolumn{1}{c|}{50} \\\cline{2-5}
       & \multicolumn{1}{c|}{\multirow{2}[2]{*}{VIX}} & \multicolumn{1}{c|}{Yes} & \multicolumn{1}{c|}{3.2415} & \multicolumn{1}{c|}{83.3} \\
       & \multicolumn{1}{c|}{} & \multicolumn{1}{c|}{No} & \multicolumn{1}{c|}{5.7268} & \multicolumn{1}{c|}{50} \\ \hline\hline
    \multirow{4}[4]{*}{Gold} & \multicolumn{1}{c|}{\multirow{2}[2]{*}{USD}} & \multicolumn{1}{c|}{Yes} & \multicolumn{1}{c|}{1.5245} & \multicolumn{1}{c|}{78.6} \\
       & \multicolumn{1}{c|}{} & \multicolumn{1}{c|}{No} & \multicolumn{1}{c|}{1.5555} & \multicolumn{1}{c|}{64.3} \\\cline{2-5}
       & \multicolumn{1}{c|}{\multirow{2}[2]{*}{VIX}} & \multicolumn{1}{c|}{Yes} & \multicolumn{1}{c|}{0.2534} & \multicolumn{1}{c|}{71.9} \\
       & \multicolumn{1}{c|}{} & \multicolumn{1}{c|}{No} & \multicolumn{1}{c|}{5.2724} & \multicolumn{1}{c|}{56.1} \\ \hline \hline
    \multirow{4}[4]{*}{Euro} & \multicolumn{1}{c|}{\multirow{2}[2]{*}{EURUSD}} & \multicolumn{1}{c|}{Yes} & \multicolumn{1}{c|}{2.6224} & \multicolumn{1}{c|}{74.1} \\
       & \multicolumn{1}{c|}{} & \multicolumn{1}{c|}{No} & \multicolumn{1}{c|}{4.3541} & \multicolumn{1}{c|}{58.6} \\\cline{2-5}
       & \multicolumn{1}{c|}{\multirow{2}[2]{*}{VIX}} & \multicolumn{1}{c|}{Yes} & \multicolumn{1}{c|}{4.4124} & \multicolumn{1}{c|}{69} \\
       & \multicolumn{1}{c|}{} & \multicolumn{1}{c|}{No} & \multicolumn{1}{c|}{4.7878} & \multicolumn{1}{c|}{53.4} \\ \hline
    \bottomrule
    \end{tabular}%
\label{tab:forecasting_results}
\end{table}%

In the dataset we have time series for the total of 66 weeks, out which we use approximately 76\% i.e. 50 weeks for the training both the models given in equation \ref{eqn:With Predictors} and \ref{eqn:Without Predictors}( for the time period 2nd June 2010 to 27th May 2011). Further we verify the model performance as one step ahead forecast over the testing period of 16 weeks from May 30th to 13 September 2011 which count for \emph{wide and robust range of market conditions}. Forecasting accuracy in the testing period is compared for both the models in each case in terms of mean absolute percentage error (MAPE) and the direction accuracy. MAPE is given by the equation \ref{eqn:mape}, where $\hat{y_i}$ is the predicted value and $y_i$ is the actual value.
\vspace{-0.09cm}
\begin{equation}
\label{eqn:mape}
 MAPE= \frac{{{\Sigma^{n}}_{i}} |\frac{y_i - \hat{y_i}}{y_i}|}{n} \times 100
\end{equation}
\vspace{-0.09cm}
While direction accuracy is measure of how accurately market or commodity up/ down movement is predicted by the model, which is technically defined as logical values for $(y_{i, \hat{t}+1}- y_{i,t})\times(y_{i,t+1}- y_{i,t})>0$ respectively. This is of prime importance to the high frequency traders and investors who hedge their investment in derivative markets as lots of prices (option premium, bonds etc.) are solely determined by the direction of the moving index or price.

As given in Table \ref{tab:forecasting_results} we observe that the there is significant reduction in the values of MAPE for all the sectors for the forecasting model with the use of predictor sentiment and SVI series than the predictor model without the use of the these predictor series.
Also for index values of DJIA direction accuracy of uptil 94.3\% is achieved, while it is for 90\% for NASDAQ-100. SVI and measure of wisdom of crowd on Twitter gives quite a robust picture of how changing dynamics of the public opinion can be reflective of the market movements that would happen in near future.

\vspace{-0.15cm}
\begin{table}[htbp]
  \centering
  \scriptsize
    \caption{Comparison with prior work in sentiment analysis for predicting markets}
      \begin{tabular}{|p{0.95cm}|p{2cm}|p{1.5cm}|p{2.1cm}|}
    \addlinespace
    \toprule
    Previous Approaches & Bollen et al. \cite{Bollen_Mao_Zeng_2010,Bollen_second_paper} and Gilbert et al. \cite{Gilbert_Karahalios_2010} &  Sprenger et al. \cite{Sprenger} & Our Approach\\ \hline
    \midrule
    Approach & Mood of complete Twitter feed & Stock Discussion with ticker \$ on Twitter & Combining Twitter sentiment + Google search volumes\\ \hline
    Dataset & 28th Feb 2008 to 19th Dec 2008, 9M tweets sampled as 1.5\% of Twitter feed & 1st Jan 2010 to 30th June 2010- 0.24M tweets & 2nd June 2010 to 13th Sept 2011- 1.9M tweets through search API\\ \hline
    Techniques & SOFNN, Grangers and linear models & OLS Regression and Correlation & Cross- Corr, GCA, Expert Model Mining System (EMMS)\\ \hline
    Results & 86.7\% directional accuracy for DJIA & Corr values uptil 0.41 for S\&P 100 stocks & Corr uptil 0.82  for OIL, DJIA, NASDAQ-100, Gold and Euro. Directional accuracy uptil 94\%\\ \hline
    Feedback/ Drawbacks & Individual modeling for stocks not feasible & News not taken into account, very less tweet volumes & Comprehensive and customizable approach \\ \hline
    \bottomrule
    \end{tabular}%
    \label{tab:comparison_table}
\end{table}%

\section {DISCUSSIONS}
From Table \ref{tab:comparison_table}, we can see that earlier works in the area of behavioral finance were limited to profile of mood states and dimensions of public mood in context of investing.
Primary objective of this work is to bring out a uniform model combining search volume behavior along with \textit{how people are speaking and about what?} on Twitter and observe how severe or accurate these effects get over the increasing time lag. Use of bullishness, agreement and message volume although non-linearly dependent on each other, provides additional features to measure sentiment in a subjective manner and also provides better understanding of variable importance. As seen in Table \ref{tab:forecasting_results}, we observe one of the most significant improvements for NASDAQ-100's VIX (MAPE- 3.2415) and Gold VIX (MAPE- 0.2534), indicating the tech savvy investors who tweet a lot, hold significant power for the index movements. Comparing general prediction performance of behavior features (SVI + Twitter sentiment series) for market indices and commodity prices over the VIX index; better performance can be observed for VIX accounting for the fact that these behavior features are better indicative of investor fear before the actual price movement occurs in the stock. However for the forex price of Euro, investor sentiment is more centralized factor in controlling the price movement as compared to the VIX index.
Modeling market sentiment is luring area that investors are looking forward for use in hedging the investment instruments.
Our results show, there is no clear uniform pattern observed in the relationships across various elements of investor behavior and stocks over a wide spectrum of market securities and indices. Hence general conclusions regarding complexities about the performance cannot be drawn. We actively look forward into the future when fully automated bots will be advanced enough to understand all the behavioral mood dimensions of location specific discussions of what an investor is saying and make successful investments at minute time-scales. There are always limitation to all predictive approaches. As statistician Box rightly said- \emph{'all models are wrong, but some are useful'}. With the limitations of natural language processing techniques, there exists multitude of problems associated with higher mood states and learning extensions to other algorithms~\cite{Alec_Bhayani}. We also observe for some cases, there is fall in the returns due to excessive rise in the bullish (positive) sentiment for a commodities/ indices. However some variation exists for product companies like EBay, Dell etc. due to excessive discussion related to product offerings and promoted messages instead of general discussion by people.

\section {CONCLUSION AND FUTURE WORK}
\label{conclusion}
Proposed approach combines the advantage of sophisticated statistical and linguistics summarization techniques. Such methods are able to capture a good picture of both the changing rates along with rise and fall probabilities for both commodities and stocks. Performance of the proposed model is more accurate in comparison to earlier works which were restricted only to mood states of entire Twitter feed applied in general to the market index~\cite{Bollen_Mao_Zeng_2010,Gilbert_Karahalios_2010,Sprenger}. We have made validation against larger tweet base, over bigger time period, with larger number of financial market instruments and greater prediction accuracy than any of the earlier works. Moreover, as far as practical implementation is concerned, our approach not only helps to improve index movements but also the present volatility and the VIX index which is the measure of the 30-days ahead market fear. More importantly it can be also used to determine portfolio adjustment decisions like ratio of risk to security for hedging with the greater confidence.

\bibliographystyle{plain}
\bibliography{bibliography}

\vspace{-0.15cm}
\section{APPENDIX}
\label{appendix}

The PCA component matrixes for Oil, DJIA, NASDAQ-100, Gold and Euro are given in Tables \ref{tab:Oil_SVI_Factors}, \ref{tab: Dow_SVI_Factors}, \ref{tab:Nasdaq_SVI_Factors},  \ref{tab:Gold_SVI_Factors} and \ref{tab:Euro_SVI_factors} respectively. Feature reduction is an important step before development of any model so as to increase predictive accuracy, simplicity and comprehensibility of the mined results. Effect of so many search terms can be concisely mapped to double or single factors i.e. original high-dimensional data onto a lower dimensional space. The new PCA factors are uncorrelated, and are ordered by the fraction of the total information each retains and filtered out on the basis of Kaiser criterion, with threshold eigen value greater than 1. Each of the search term factors (Fact 1 and Fact 2) explain significant amount of variance as in the original feature set of search keywords given in the Tables below.

\begin{table}[htbp]
  \centering
 \caption{Vector Matrix for Gold SVI Factors (\emph{Only 1 factor for gold as search terms for gold fall on the same dimension plane in the feature vector map.}) }
    \begin{tabular}{|l|c|}
    \toprule
    \multicolumn{1}{|c|}{\multirow{2}[4]{*}{Search Term}} & \multicolumn{1}{|p{2.5cm}|}{Gold SVI factors} \\
    \multicolumn{1}{|c|}{} & \multicolumn{1}{|c|}{Fact 1} \\  \midrule  \hline
    \multicolumn{1}{|l|}{buy gold} & .575 \\
    \multicolumn{1}{|l|}{invest in gold US data} & .942 \\
    \multicolumn{1}{|l|}{invest in gold worldwide} & .884 \\
    \multicolumn{1}{|l|}{dollar to pound exchange rate} & .905 \\
    \multicolumn{1}{|l|}{dollar to pound exchange} & .904 \\  \hline
    \bottomrule

    \end{tabular}%
\vspace{-0.2cm}
\label{tab:Gold_SVI_Factors}
\end{table}%
\vspace{-0.15cm}

\begin{table}[htbp]
  \centering
  \caption{Vector Matrix for Oil SVI Factors}
    \begin{tabular}{|c|p{2cm}|p{2cm}|}
    \toprule
    \multicolumn{1}{|c|}{\multirow{2}[4]{*}{Search Terms}} & \multicolumn{2}{|c|}{Oil SVI Factors} \\
    \multicolumn{1}{|c|}{} & \multicolumn{1}{|p{1.5cm}|}{Fact 1} & \multicolumn{1}{|p{1.5cm}|}{Fact 2} \\ \midrule  \hline
    \multicolumn{1}{|l|}{oil commodity} & .671 & .688  \\
    \multicolumn{1}{|l|}{crude oil etf} & .319 & .897 \\
    \multicolumn{1}{|l|}{oil funds} & .304 & .900  \\
    \multicolumn{1}{|l|}{oil etf} & .303 & .898  \\
    \multicolumn{1}{|l|}{oil quotes} & .593 & .599 \\
    \multicolumn{1}{|l|}{oil prices per barrel} & .853 & .338  \\
    \multicolumn{1}{|l|}{spot oil prices} & .727 & .533 \\
    \multicolumn{1}{|l|}{wti crude} & .361 & .504  \\
    \multicolumn{1}{|l|}{how much oil is left } & .452 & .446\\
    \multicolumn{1}{|l|}{futures price} & .521 & .759\\
    \multicolumn{1}{|l|}{how to buy oil} & .298 & .918\\
    \multicolumn{1}{|l|}{oil ticker} & .634 & .655 \\
    \multicolumn{1}{|l|}{current oil} & .741 & .021 \\
    \multicolumn{1}{|l|}{crude oil futures} & .441 & .639 \\
    \multicolumn{1}{|l|}{crude oil price} & .442 & .700 \\  \hline
    \bottomrule
    \end{tabular}%
     \vspace{-0.2cm}
\label{tab:Oil_SVI_Factors}
\end{table}%

\vspace{-0.15cm}
\begin{table}[htbp]
  \centering
  \caption{Vector Matrix for DJIA SVI Factors}
    \begin{tabular}{|l|p{1cm}|p{1cm}|}
    \toprule
    \multicolumn{1}{|c|}{\multirow{2}[4]{*}{Search Terms}} & \multicolumn{2}{|c|}{Dow SVI Factors} \\
    \multicolumn{1}{|c|}{} & \multicolumn{1}{|p{1.5cm}|}{Fact 1} & \multicolumn{1}{|p{1.5cm}|}{Fact 2} \\  \midrule \hline
    \multicolumn{1}{|l|}{djia} & .931  & .038 \\
    \multicolumn{1}{|l|}{dow jones industrial average} & .811  & .495 \\
    \multicolumn{1}{|l|}{dow jones} & .929  & .116 \\
    \multicolumn{1}{|l|}{dow} & .966  & .010 \\
    \multicolumn{1}{|l|}{s\&p 500} & .635  & -.032 \\
    \multicolumn{1}{|l|}{Stock Market} & .689  & -.337 \\
    \multicolumn{1}{|l|}{stock message board} & -.012 & .936 \\  \hline
    \bottomrule
    \end{tabular}%
     \vspace{-0.2cm}
\label{tab: Dow_SVI_Factors}
\end{table}%

\vspace{-0.15cm}
\begin{table}[htbp]
  \centering
  \caption{Vector Matrix for Nasdaq SVI Factors }
    \begin{tabular}{|l|p{1cm}|p{1cm}|}
    \toprule
    \multicolumn{1}{|c|}{\multirow{2}[4]{*}{Search Terms}} & \multicolumn{2}{|c|}{Nasdaq SVI Factors}\\

    \multicolumn{1}{|c|}{} & \multicolumn{1}{|p{1.5cm}|}{Fact 1} & \multicolumn{1}{|p{1.5cm}|}{Fact 2} \\ \midrule  \hline
    \multicolumn{1}{|l|}{nasdaq today} & .833 & .398 \\
    \multicolumn{1}{|l|}{dow futures quote} & .888 & .239 \\
    \multicolumn{1}{|l|}{futures quote} & .673 & -.004 \\
    \multicolumn{1}{|l|}{NASDAQ quote} & .821 & .326 \\
    \multicolumn{1}{|l|}{nylc} & .257 & .936 \\
    \multicolumn{1}{|l|}{bank of america dividends} & .161 & .947 \\  \hline
    \bottomrule
    \end{tabular}%
     \vspace{-0.2cm}
\label{tab:Nasdaq_SVI_Factors}
\end{table}%

\vspace{-0.15cm}

\begin{table}[htbp]
  \centering
 \caption{Vector Matrix for Euro SVI Factors}
    \begin{tabular}{|p{1cm}|p{1cm}|p{1.5cm}|}
    \toprule
    \multicolumn{1}{|p{1cm}|}{\multirow{2}[3]{*}{Search Terms}} & \multicolumn{2}{|c|}{Euro SVI factors}\\

    \multicolumn{1}{|p{1cm}|}{} & \multicolumn{1}{|p{1.5cm}|}{Fact 1} & \multicolumn{1}{|p{1.5cm}|}{Fact 2} \\  \midrule \hline
    \multicolumn{1}{|l|}{exchange rates converter} & .530 & -.383 \\
    \multicolumn{1}{|l|}{dollar euro exchange rate history} & .798 & .079  \\
    \multicolumn{1}{|l|}{rupee exchange rate} & -.047 & .063\\
    \multicolumn{1}{|l|}{oanda currency} & .105 & .810 \\
    \multicolumn{1}{|l|}{rupee exchange} & .929 & -.102 \\
    \multicolumn{1}{|l|}{dollar rupee exchange rate} & .938 & .043 \\
    \multicolumn{1}{|l|}{bloomberg live tv} & .828 & -.228 \\
    \multicolumn{1}{|l|}{eurusd} & -.237 & .741 \\ \hline
    \bottomrule
    \end{tabular}
     \vspace{-0.2cm}
\label{tab:Euro_SVI_factors}
\end{table}
\balancecolumns

\end{document}